\documentclass{PoS}
\usepackage{bbold}

\def\eqref#1{{(\ref{#1})}}
\def\mc#1{{\mathcal #1}}
\def\mbf#1{{\mathbf #1}}

\def\d{\delta}
\def\D{\Delta}
\def\g{\gamma}
\def\L{\Lambda}
\def\O{\Omega}
\def\S{\Sigma}

\title{Nuclear Physics Review}

\ShortTitle{Nuclear Physics Review}

\author{\speaker{Andr\'{e} Walker-Loud}\thanks{New affiliation: The College of William and Mary and Jefferson Laboratory}\\
        Lawrence Berkeley National Laboratory, Berkeley, CA 94720, USA\\
        Department of Physics, University of California, Berkeley, CA 94720, USA \\
         E-mail: \email{walkloud@wm.edu}}

\abstract{Anchoring low-energy nuclear physics to the fundamental theory of strong interactions remains an outstanding challenge.
I review the current progress and challenges of the endeavor to use lattice QCD to bridge this connection.
This is a particularly exciting time for this line of research as demonstrated by the spike in the number of different collaborative efforts focussed on this problem and presented at this conference.
I first digress and discuss the 2013 Ken Wilson Award.
}

\FullConference{31st International Symposium on Lattice Field Theory - LATTICE 2013\\
		July 29 - August 3, 2013\\
		Mainz, Germany}

\begin{document}

\section{Kenneth G. Wilson \label{sec:KW}}
It is a great honor to receive this award~\cite{kw_award} and simultaneously bittersweet, given the recent passing of Ken Wilson~\cite{Kronfeld:2013lda}.
It leaves with me with a great confusing mix of feelings and thoughts which I will mostly spare you.
There are two thoughts I feel compelled to share: I would have liked to meet him; I feel a great sense of responsibility to continue doing good research, and to strive to make more significant contributions, to both live up to this award and to give back to our field.

Amongst all those who have had an influence on my physics education and development, there are a few in particular I would like to acknowledge: Martin Savage, my Ph.D advisor who helped me learn to motivate myself through some mix of fear and high standards; Steve Sharpe, my pseudo-Ph.D advisor who happily tolerated many unwarned conversations about physics and life; Will Detmold, David Lin and Brian Tiburzi, whose doors were always open to my near endless list of physics questions both during my graduate studies and after; Paulo Bedaque and Kostas Orginos who both helped me grow into a real scientist through sound advice whether followed or not; Maarten Golterman who has entertained many of my more bizarre questions and helped clarify many subtle physics points; Wick Haxton who has been a fantastic mentor and helped me to sharpen my research focus; and of course, all my other research collaborators and friends.
I would like to give an especial acknowledgement to all those who have made contributions to Lattice Field Theory as significant or more so than my own, who were not eligible simply because of the rules of consideration, particularly the other ``young'' researchers as deserving as myself.

I was asked to give a short presentation based upon the work recognized by this award, received \textit{For significant contributions to our understanding of baryons using lattice QCD and effective field theory}.

Effective field theory (EFT) teaches us how things should be \dots I grew up learning effective field theory.
Lattice QCD (LQCD) teaches us how things are \dots in my postdoc youth, I learned some lattice QCD.
Lattice QCD provides numerical answers to specific questions.  EFT provides a framework to understand these numbers in a broader context, and provides a quantitative connection with many other questions.
Chiral Perturbation Theory ($\chi$PT), the low-energy EFT of QCD, has been needed to extrapolate results from LQCD calculations to the real world: the physical quark masses, the infinite volume, and assisting in the continuum limit extrapolation.
Lattice QCD calculations are now performed close to the real world: LQCD can now be used to significantly improve our understanding of $\chi$PT.

What separates $\chi$PT from a simple Taylor expansion?  The chiral expansion informs you approximately the range of validity of the theory (EFT).  $\chi$PT is described by universal coefficients which describe many observables.  $\chi$PT predicts \textit{chiral logarithms} or rather non-analytic dependence upon the light-quark masses which arise from long-range pion physics that can not necessarily be well modeled with a Taylor (local) expansion.
Evidence for these chiral logarithms is deemed essential for finding the \textit{chiral regime} with sufficiently light quark masses that the chiral expansion is likely converging.

\begin{figure}
\center
\includegraphics[width=0.38\textwidth]{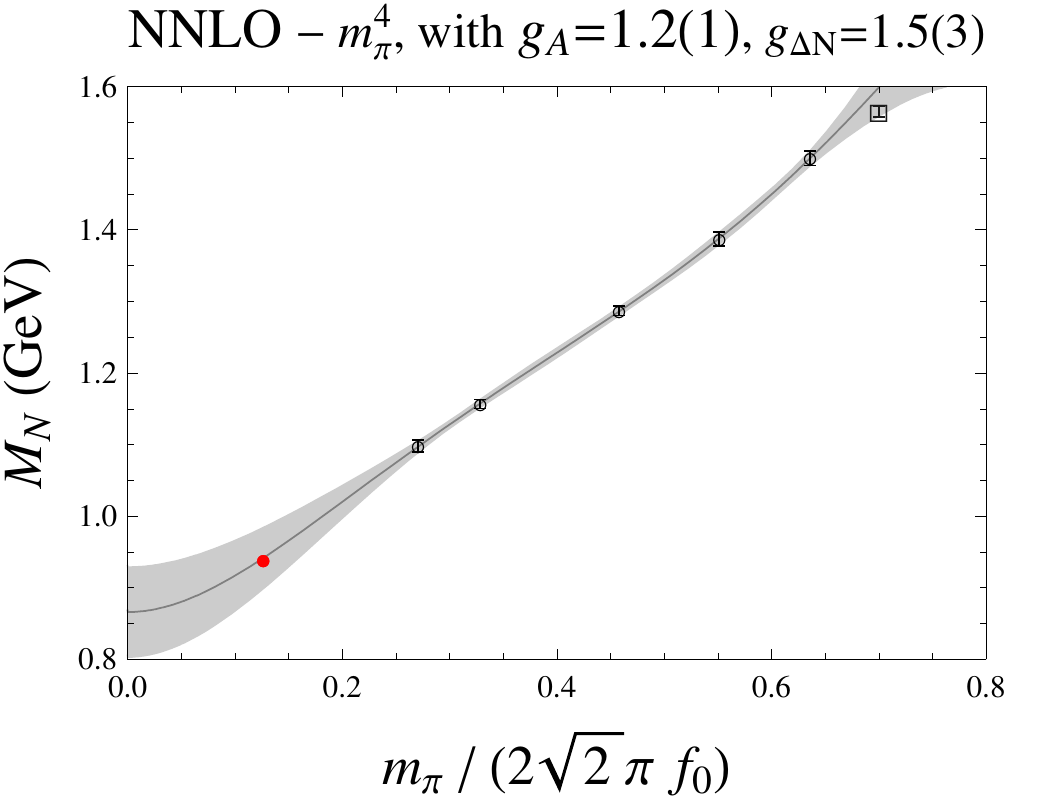}
\includegraphics[width=0.44\textwidth]{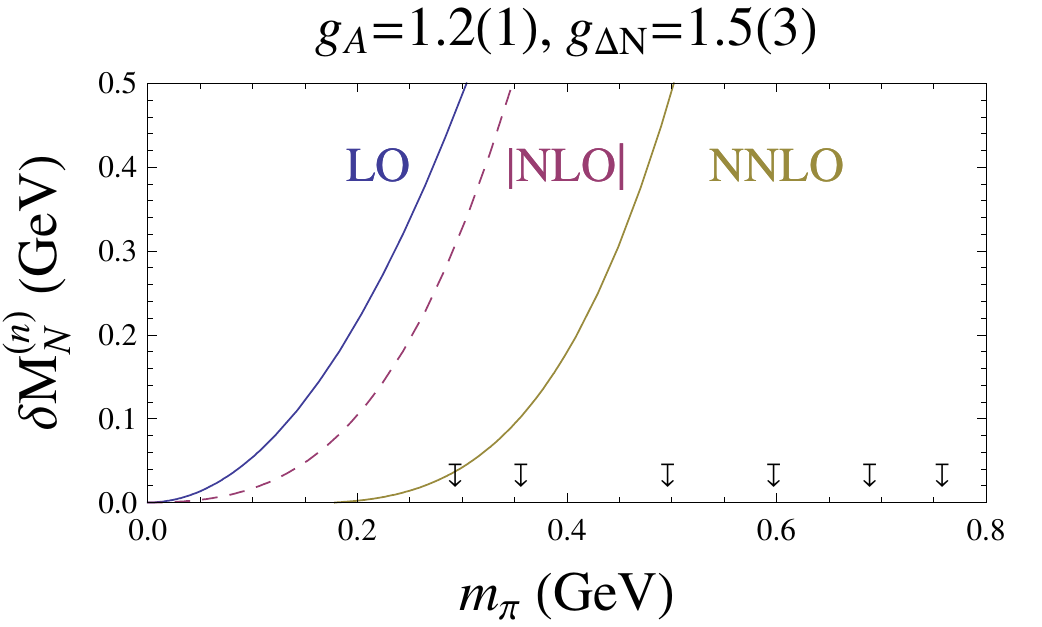}
\caption{\label{fig:mn_lhp}The fit (left) and convergence (right) of the NNLO nucleon mass formula compared with the LHPC results~\cite{WalkerLoud:2008bp}.  In the convergence plot, the arrows indicate the values of the pion masses used.  The resulting fit does not include the physical point denoted by a red circle.  The pion mass has been scaled by $\Lambda_{\chi_0} = 2\sqrt{2} \pi f_0$ where $f_0$ is the pion decay constant in the chiral limit, such that the x-axis is approximately the chiral expansion parameter for baryon $\chi$PT.}
\end{figure}

\begin{figure}
\center
\includegraphics[width=0.4\textwidth]{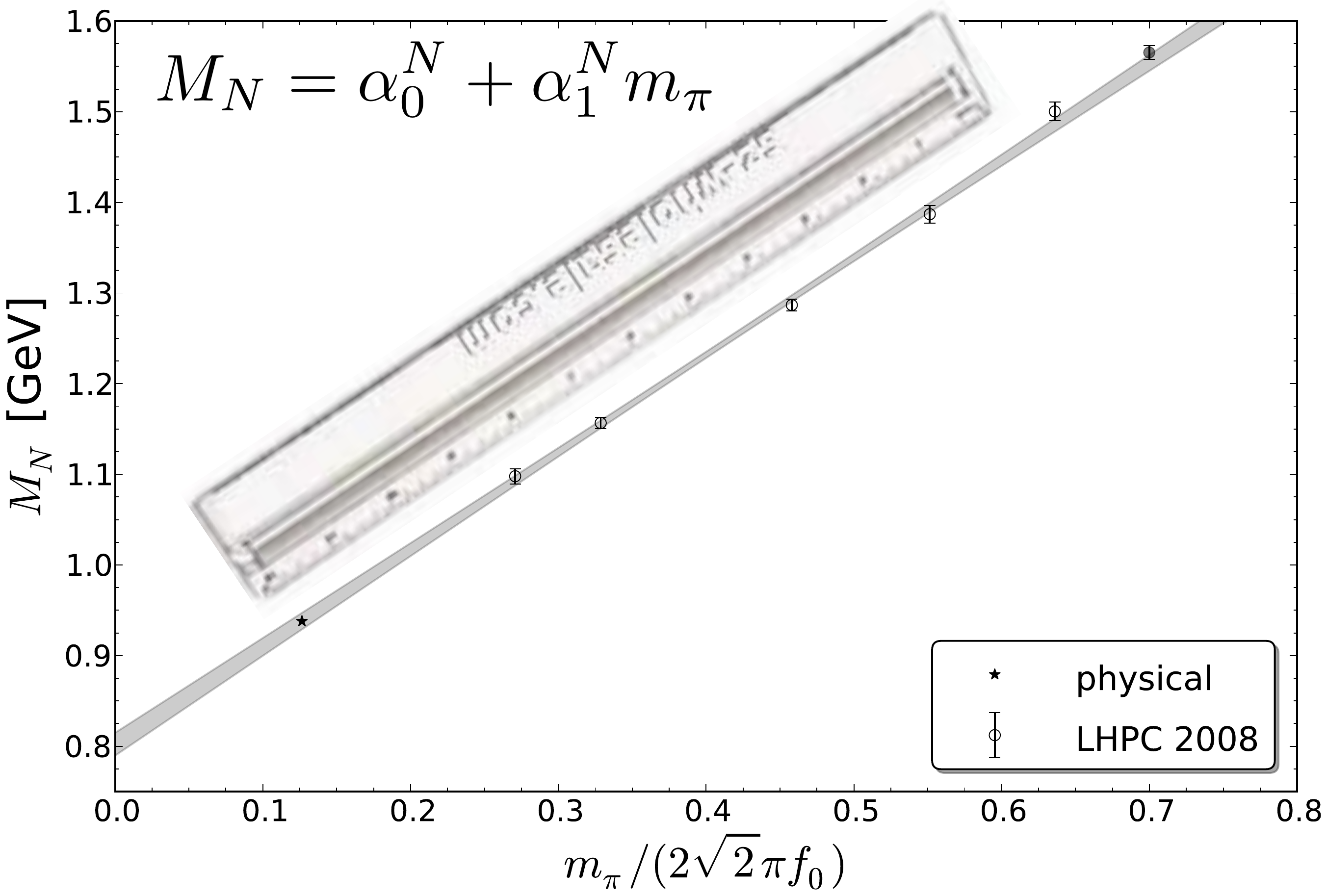}
\includegraphics[width=0.4\textwidth]{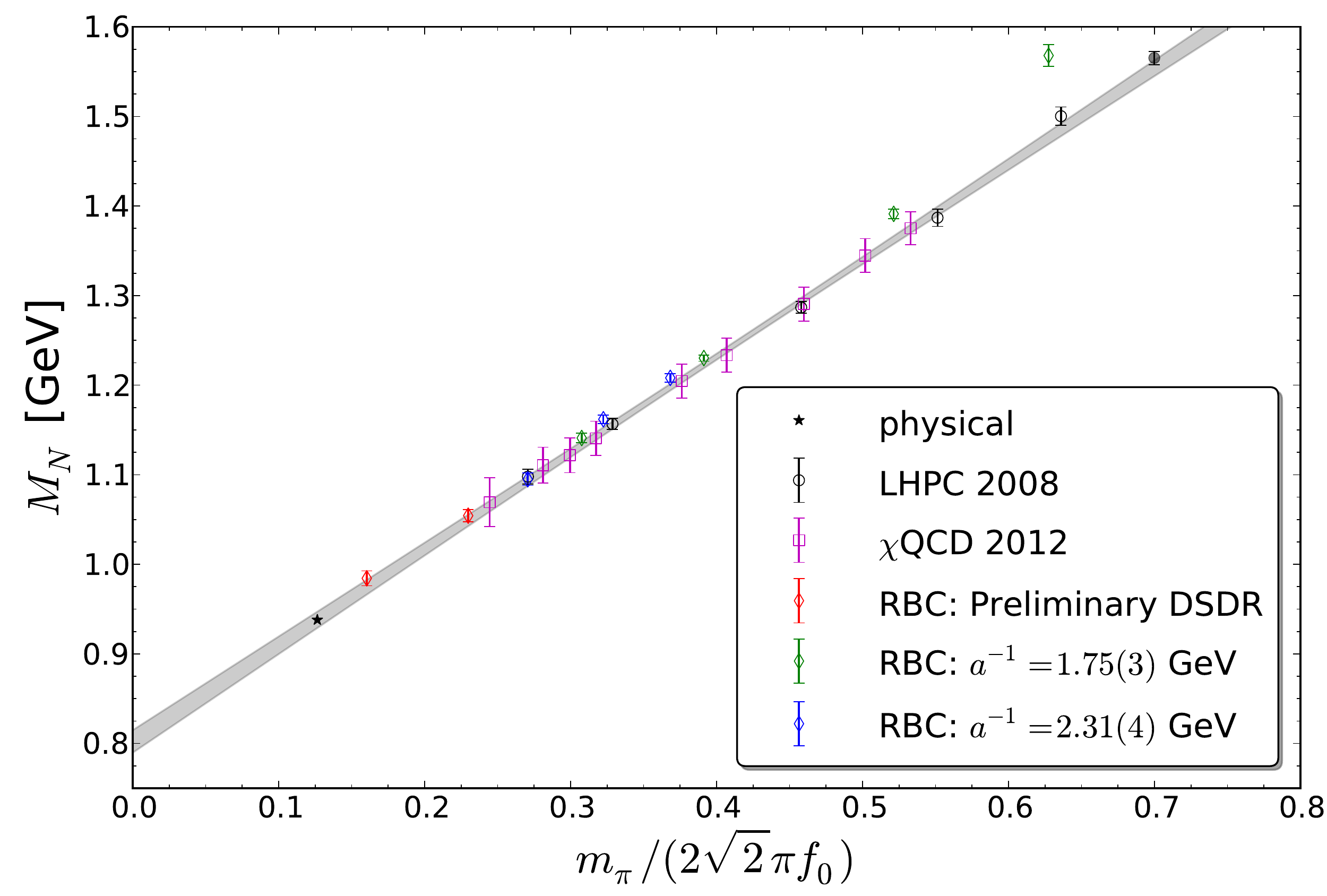}
\caption{\label{fig:mn_ruler}The ``ruler'' plot (left).  Within uncertainties, $M_N[\textrm{MeV}] = 800 + m_\pi$.
Updated results (right) presented at Chiral Dynamics 2012~\cite{Walker-Loud:2013yua}, including those published in \cite{Gong:2013vja}.}
\end{figure}

In the chiral expansion, all hadron masses can be expanded in a quark mass expansion 
\begin{equation}
	M_H = M_{H,0} + \alpha_H m_l + \dots
\end{equation}
where $M_{H,0}$ is the hadron mass in the chiral limit and $m_l$ is the light quark ($u,d$) mass.
The exception to this rule are the pions (kaons, $\eta$) whose masses vanishes in the chiral limit as they are the pseudo Nambu-Goldstone bosons arising from the spontaneous breaking of the global chiral symmetry in the QCD action,
\begin{equation}
m_\pi^2 = -2 m_l \frac{\langle 0 | \bar{q}_l q_l | 0 \rangle}{f^2}
+\dots
\end{equation}
For example, the nucleon mass is given at next-to-leading order (NLO) in $\chi$PT 
\begin{equation}
\label{eq:mn_nlo}
M_N = M_{N,0}(\mu) + \frac{\alpha_N(\mu)}{4\pi f_\pi} m_\pi^2
	-\frac{3\pi g_A^2}{(4\pi f_\pi)^2} m_\pi^3
	-\frac{8g_{\pi N \Delta}^2}{3(4\pi f_\pi)^2} \mathcal{F}(m_\pi,\Delta,\mu)
	+\dots
\end{equation}
We see above that the non-analytic terms, $m_\pi^3, \mathcal{F}(m_\pi,\Delta,\mu)$ arise only at NLO.  Can we observe this non-analytic light-quark mass dependence in the numerical results of the nucleon spectrum?
This is a question I looked at in detail with the LHP Collaboration~\cite{WalkerLoud:2008bp} and expanded upon at the 2008 Lattice conference~\cite{WalkerLoud:2008pj}.
We found the NLO formula was insufficient to describe the numerical results if one demanded the nucleon axial charge (and $g_{\pi N\Delta}$) be close to its physical value as these mass corrections are strictly negative while the numerical results of the nucleon mass increase with increasing quark masses.
To stabilize the fit, either one needed $g_A \sim 0$ or to include the next-to-next-to leading order (NNLO) corrections.
The full NNLO fit resulted in a good $\chi^2/dof$, an extrapolated nucleon mass in agreement with experiment, $M_N = 941 \pm 42 \pm 17$~MeV, but the convergence of the chiral expansion was marginally acceptable for the lightest pion mass ($m_\pi \sim 300$~MeV) and worse/non-convergent for the heavier points, see Fig.~\ref{fig:mn_lhp}.
More striking was the linear nature of the results plotted versus the pion mass, displayed in Fig.~\ref{fig:mn_ruler} (left) which Brian Tiburzi has coined the ``ruler'' plot.  The nucleon mass is well described by a fit form
\begin{equation}
M_N = \alpha_0^N + \alpha_1^N m_\pi\, .
\end{equation}
It was found that all LQCD calculations of the nucleon mass with $2+1$ dynamical fermions displayed this striking linear behavior~\cite{WalkerLoud:2008pj}.
I am not advocating this as a good model of QCD.
Taking this result seriously, the nucleon mass, within uncertainties can be parameterized as
\begin{equation}
	M_N[\textrm{MeV}] = 800 + m_\pi\, .
\end{equation}
\textit{I am not advocating this as a good model for QCD}.
It clearly parameterizes the numerical results in the range of available masses and agrees with the physical point.  However, it is clearly incorrect at and near the chiral limit, as it predicts the wrong quark mass dependence.

What is the status now?  At the 2012 Chiral Dynamics Workshop, I presented updated results from the RBC/UK-QCD and $\chi$QCD~\cite{Gong:2013vja} Collaborations.
The results are displayed in Fig.~\ref{fig:mn_ruler} (right) along with the original fit from LHPC~\cite{WalkerLoud:2008bp}.
There continues to be more evidence that this striking linear in $m_\pi$ dependence of $M_N$ is a feature of QCD and not a conspiracy of lattice artifacts.
This has important implications beyond mere academic curiosity.
As discussed in the recent review at Lattice 2012~\cite{Young:2013nn}, the Feynman-Hellman Theorem is one of two main methods of determining the scalar light and strange quark content of the nucleon, which utilizes the quark mass dependence of the nucleon.  For example, the light quark mass dependence is related to
\begin{equation}
	\sigma_{\pi N} \equiv
	m_l \langle N | \bar{q}_l q_l | N \rangle = 
	m_l \frac{\partial}{\partial m_l} M_N(m_l)
	\simeq
	\frac{m_\pi}{2} \frac{\partial}{\partial m_\pi} M_N(m_\pi)\, .
\end{equation}
Using the ruler approximation, one finds $\sigma_{\pi N} = 67 \pm 4$~MeV.
These scalar matrix elements have important implications for understanding direct dark matter detection experiments, as described for example in Refs.~\cite{dm}.

\subsection{Quantitatively connecting the Quarks with Big Bang Nucleosynthesis}
I would like to take this opportunity to share with you new preliminary work, of a similar vein.
I will focus on the isospin breaking of the nucleon mass, which is known very precisely experimentally~\cite{Mohr:2008fa}
\begin{equation}
	M_n - M_p = 1.29333217(42) \textrm{ MeV}\, .
\end{equation}
The Standard Model has two sources of isospin breaking
\begin{equation}
Q = \frac{1}{6} \mathbb{1} + \frac{1}{2} \tau_3\, , \qquad
m_q = \hat{m} \mathbb{1} - \delta \tau_3\, .
\end{equation}
Given only electrostatic forces, one would predict $M_p > M_n$, but we now know the contribution from $m_d - m_u$ is comparable in size but opposite in sign making the neutron slightly heavier.

This nucleon mass splitting plays an extremely significant role in the evolution of the universe as we know it.
It controls the initial conditions for Big Bang Nucleosynthesis (BBN) which describes the production of light nuclei in the early Universe.  
After the Universe cools off such that the weak interactions decouple from the expansion, the nucleons are in approximate thermodynamic equilibrium, so the ratio of neutrons to protons is given approximately by
\begin{equation}
	\frac{X_n}{X_p} = e^{- \frac{M_n - M_p}{T}}\, .
\end{equation}
Further, the neutron lifetime (and other $np$ reactions) are highly sensitive to the value of this mass splitting.  The neutron lifetime is given by
\begin{equation}
\frac{1}{\tau_n} =
	\frac{(G_F \cos \theta_C)^2}{2\pi^3} m_e^5 (1 + 3g_A^2)
	f \left( \frac{M_n - M_p}{m_e} \right)\, ,
\end{equation}
where $f(q)$ is a function of the decay phase-space.
Approximating the nucleons as point particles yields~\cite{Griffiths:2008zz} 
$f(q) = \frac{1}{15}(2q^4 -9q^2 -8) \sqrt{q^2 -1} +q \ln(q + \sqrt{q^2 -1} )$; a 10\% change in the nucleon mass splitting results in a $\sim$100\% change in the neutron lifetime.
How does a change in $M_n - M_p$ then propagate into BBN?

BBN describes the production of light nuclei through a set of coupled nuclear reactions.
Given the measured reactions, the only input/output to our understanding is the primordial baryon to photon ratio, $\eta = X_N / X_\gamma$.  This quantity is now known precisely also from the Cosmic Microwave Background and measured to be $\eta = 6.23(17)\times 10^{-10}$~\cite{Dunkley:2008ie}, in excellent agreement with the predicted value from BBN.
A good review of BBN can be found in Ref.~\cite{Olive:2010mh}.

During this epoch, there are a few important time scales set by basic nuclear physics.
About one second after the Big Bang, or when the temperature is $\sim 1$~MeV, the Universe is composed of protons, neutrons, electrons, photons and neutrinos.  
The reaction $n+p \leftrightarrow d + \gamma$ occurs roughly equally in both directions until the Universe expands and cools off to a temperature of $T\sim 0.1$~MeV which occurs roughly 3 minutes after the Big Bang.
At this time, the ``deuterium bottleneck'' is surpassed and the Universe rapidly forms deuterium and ${}^4$He.
Why does the Universe not form deuterium earlier as the binding energy is $B_d \simeq 2.2$~MeV?
This is because of the approximately one billion photons for every nucleon $(\eta)$, so the long Boltzmann tail of the photon gas keeps dissociating deuterium as quickly as it is formed until the Universe cools sufficiently.  The precise time is sensitive to $B_d$ which is a finely tuned quantity in nature.
After the formation of ${}^4$He, trace amounts of other nuclei are formed but the lack of bound $A=5$ or $A=8$ nuclei limit their formation in the early Universe.  
After $\tau_n\sim 15$~minutes, the remaining free neutrons decay leaving a primordial Universe composed of $\sim$75\% Hydrogen, $\sim$25\% ${}^4$He and trace amounts of other light nuclei by mass fraction.
How would this picture change with a different value of $M_n - M_p$?  A larger isospin mass splitting means the neutrons decay more quickly, leaving more Hydrogen, and hence more stars like our Sun, while a smaller isospin splitting leads to a neutron rich Universe.

We would like to understand $M_n -M_p$ directly from first principles.
At leading order (LO) in isospin breaking, the nucleon mass splitting can be separated into two corrections
\begin{equation}
	\d M_N \equiv M_n - M_p = 
	\d M^\g + \d M^{m_d - m_u}\, .
\end{equation}
The disparate length scales relevant to QCD and QED make precise LQCD calculations of the electro-magnetic self-energy, $\d M^\g$ challenging, while the strong isospin breaking correction is perfectly suited to lattice calculations~\cite{lattice_isospin}.
There is an alternate means of computing $\d M^\g$ using the Cottingham Formula~\cite{Cottingham} which relates the electromagnetic self-energy to forward Compton scattering through dispersion integrals.
This determination of $\d M^\g$ was updated recently with our modern knowledge of nucleon structure~\cite{WalkerLoud:2012bg}
\begin{equation}
\label{eq:cottingham}
	\d M^\g_{p-n} = 1.30 \pm 0.03 \pm 0.47 \textrm{ MeV}\, ,
\end{equation}
where the first uncertainty is propagated from the measured uncertainty of nucleon structure, and the second uncertainty arises from an unavoidable and unknown subtraction function arising in the Cottingham formulation.  Formally, the subtraction function is known exactly in the low and high $Q^2$ regions so progress can be made in its parameterization through improved knowledge of the nucleon polarizabilities and other low-energy nucleon structure.  The large uncertainty presently comes from our lack of constraint on the iso-vector nucleon magnetic polarizability~\cite{Griesshammer:2012we} which is being addressed with LQCD~\cite{emc:pols}.

\begin{figure}
\center
\includegraphics[width=0.5\textwidth]{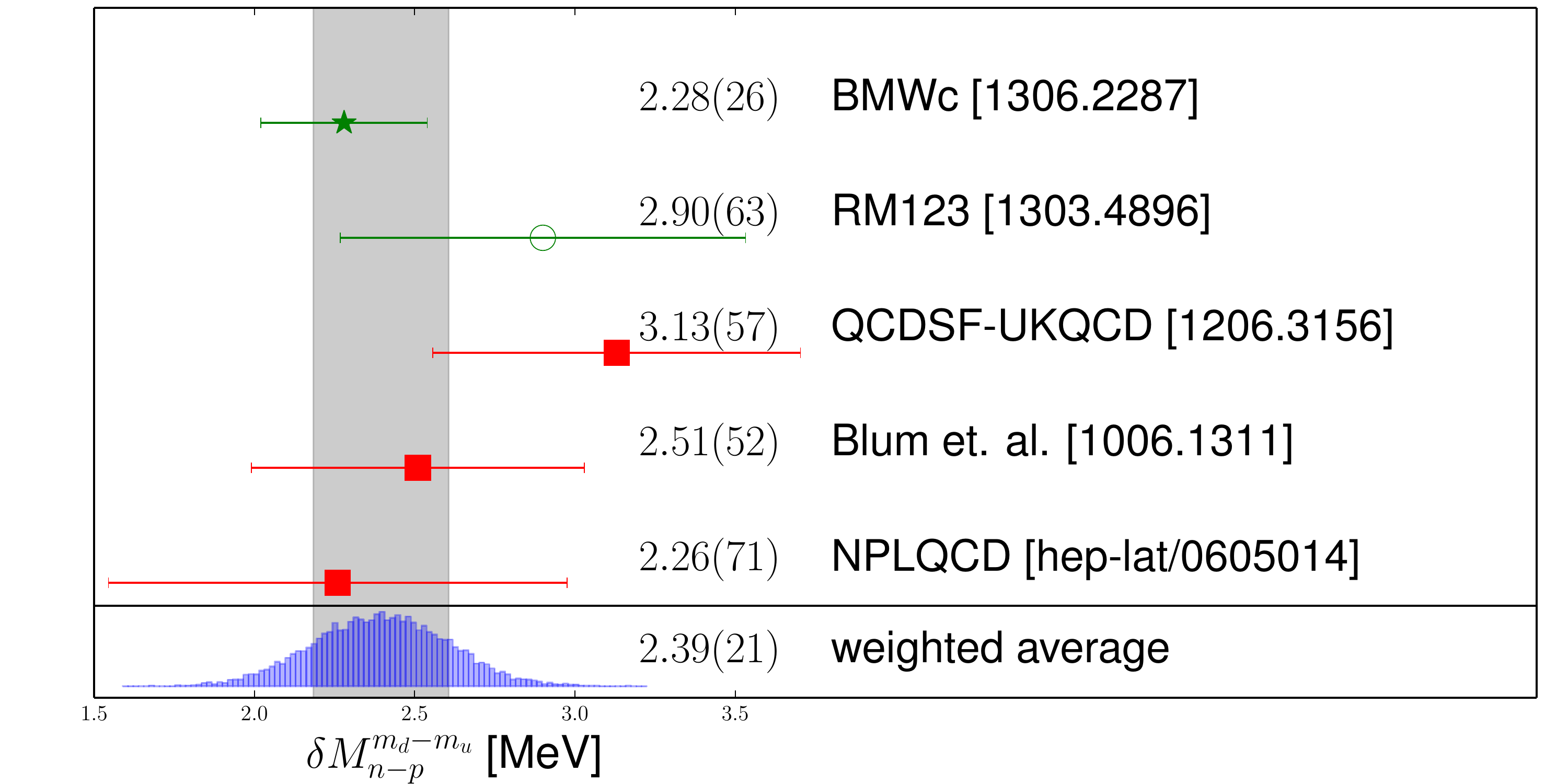}
\caption{\label{fig:mn-mp}Current LQCD calculations of $\d M_{n-p}^{m_d-m_u}$ with a color scheme similar to FLAG~\cite{Aoki:2013ldr}.}
\end{figure}

There are now several lattice calculations of $\d M_{n-p}^{m_d-m_u}$, summarized in Fig.~\ref{fig:mn-mp}.
The calculations denoted with a red square use a single lattice spacing while the green circle and star use two and five lattice spacings respectively.
Unlike in Ref.~\cite{Junnarkar:2013ac}, a simple weighted average of these results produces a seemingly reasonable lattice average.  There are still unfortunately, a small number of results, so in this average, I do not discriminate and include all results, but penalizing the red-square results as described in \cite{Junnarkar:2013ac}, arriving at $\d M_{n-p}^{m_d -m_u} = 2.39(21)$~MeV.
We can combine this result with the experimental splitting to predict the electromagnetic contribution
\begin{equation}
	\d M^\g_{p-n} = M_p - M_n -\d M_{p-n}^{m_d-m_u}
	= 1.10\pm 0.21 \textrm{ MeV}\, ,
\end{equation}
in good agreement with the estimate~\eqref{eq:cottingham}.

The lattice calculations can be improved by using a symmetric breaking of isospin of the valence quarks about degenerate sea quarks~\cite{WalkerLoud:2009nf}%
\footnote{Similar ideas were also developed and implemented by the RM123 Collaboration~\cite{lattice_isospin}.}
\begin{equation}
\label{eq:symm_isospin}
	m_l = m_{u,d}^{sea}\, , \quad
	m_u^{val} = m_l - \d\, , \quad
	m_d^{val} = m_l + \d\, .
\end{equation}
This introduces a partial-quenching (PQ) error in the calculations which scales as $\mc{O}(\d^2)$ for isospin symmetric quantities and $\mc{O}(\d^3)$ for iso-vector quantities.
These PQ effects can be understood also in PQ $\chi$PT (PQ$\chi$PT).  For example the resulting pion masses are determined at NLO
\begin{eqnarray}
m_{\pi^\pm}^2 &=& 2Bm_l \left\{
	1 + \frac{m_\pi^2}{(4\pi f_\pi)^2} \ln \left( \frac{m_\pi^2}{\mu^2} \right)
	+\frac{4m_\pi^2}{f_\pi^2} l_4^r(\mu) \right\}
	-\frac{\D_{PQ}^4}{2(4\pi f_\pi)^2}\, ,
\\
m_{\pi^0}^2 &=& m_{\pi^\pm}^2
	+\frac{16B^2 \d^2}{f_\pi^2} l_7\, .
\end{eqnarray}
In this equation, $\Delta_{PQ}^2 = 2B\d$; the isospin breaking mass term also controls the PQ effects.
Those familiar with PQ$\chi$PT will notice the lack of an enhanced chiral log~\cite{Sharpe:1997by}.
The improved chiral behavior arises specifically from this symmetric breaking of isospin, Eq.~\eqref{eq:symm_isospin}.
There is also a significant improvement to the chiral behavior of the nucleon mass splitting;
\begin{equation}
\label{eq:dmN_nnlo}
\d M_{n-p}^\d = \d \left\{ 
	2\alpha \left[ 1 
		-\frac{m_\pi^2}{(4\pi f_\pi)^2} (6g_A^2 +1) \ln \left( \frac{m_\pi^2}{\mu^2} \right)
		\right]
	+\beta(\mu) \frac{2m_\pi^2}{(4\pi f_\pi)^2}
	\right\}
	+\frac{\alpha \d \D_{PQ}^4\left(4 -3g_0^2 \right)}{m_\pi^2(4\pi f_\pi)^2}\, .
\end{equation}
In this expression, the terms in curly braces are those from $\chi$PT (QCD) while the last term proportional to $\d^3$ is from the partial quenching (for simplicity, I have ignored the coupling to the deltas in this expression).
In terms of the chiral expansion, most importantly, the problematic leading non-analytic terms ($m_\pi^3$) exactly cancel in the isospin splitting.  This cancellation only happens with symmetric isospin breaking, Eq.~\eqref{eq:symm_isospin}.
The chiral expansion of Eq.~\eqref{eq:dmN_nnlo} is then as well behaved as the chiral expansion of the pion mass or decay constant.

\begin{figure}
\center
\includegraphics[width=0.42\textwidth]{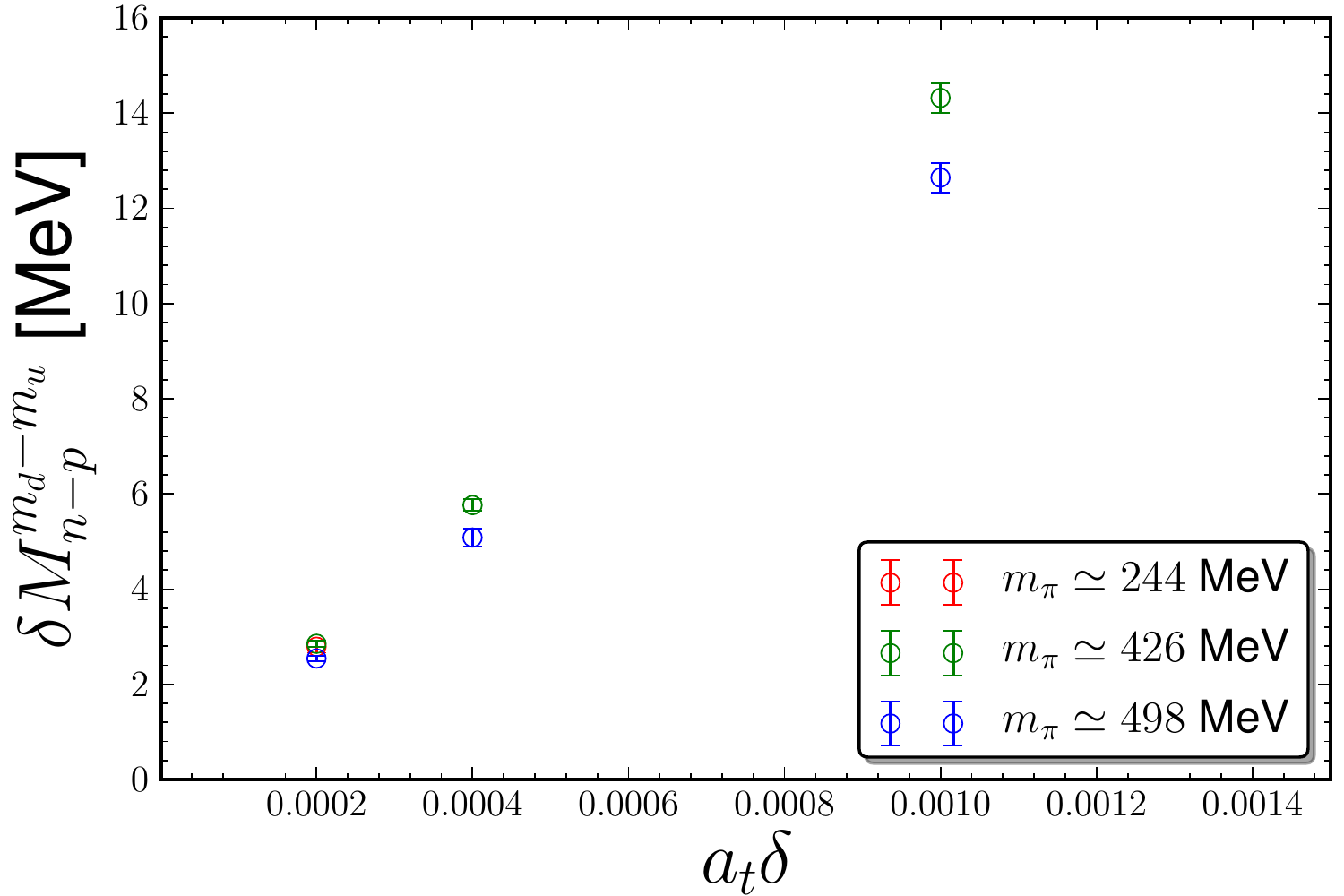}
\includegraphics[width=0.42\textwidth]{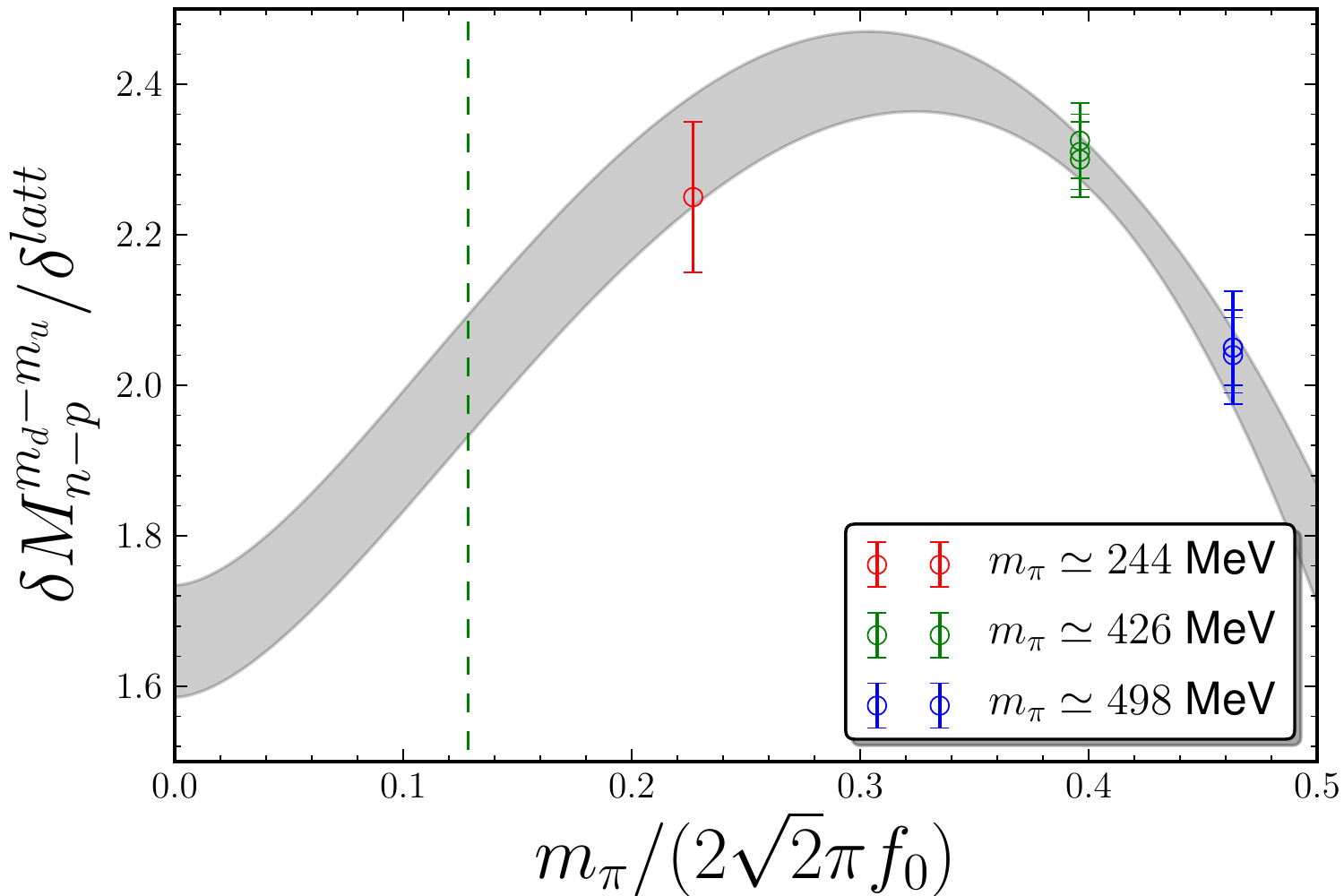}
\caption{\label{fig:emc_spec}Results~\cite{emc_spec} for $\d M_{n-p}^{m_d-m_u}$ in physical units vs the lattice value of $a_t \d$ (left) and also $\d M_{n-p}^{m_d-m_u}/\d$ vs the pion mass (right).
The lightest pion mass point has a single value of $\d$ while the heavier two points have 3 values of $\d$.  The vertical dashed line is at $m_\pi^{phy}$.}
\end{figure}

How does this predicted pion mass dependance \eqref{eq:dmN_nnlo} compare with the numerical lattice results?
Here, I report preliminary results using the anisotropic clover-Wilson ensembles produced by HSC~\cite{Lin:2008pr}.
Our results use three different values of $\d$ and three values of the light quark mass corresponding to $m_\pi \simeq \{244,426,498\}$~MeV ($M_\O$ scale setting)~\cite{emc_spec}.
The results are displayed in Fig.~\ref{fig:emc_spec} along with the resulting fit (gray band) utilizing Eq.~\eqref{eq:dmN_nnlo}.
If the nucleon axial charge is put to its physical value, $g_A = 1.27$, the resulting $\chi^2/dof = 1.34/4 = 0.33$ is quite good.
If the axial coupling is left as a free parameter, it is determined in the minimization to be $g_A^{fit} = 1.5(.3)$, in very good agreement with the physical value.
This is to be contrasted with a fit to the nucleon mass which returns $g_A \sim 0$, mentioned above.
Removing the heaviest pion mass results from the fit yields indistinguishable results except with larger uncertainties.
The strong curvature observed in the results is due to the chiral logarithm in Eq.~\eqref{eq:dmN_nnlo}, with, I note, a particularly large pre-factor, $6g_A^2 +1$.
\textit{Taken all together, this is striking evidence of non-analytic light quark mass dependence in the nucleon spectrum.}%
\footnote{I have previously reported on evidence for such non-analytic light quark mass dependence looking at octet-decuplet mass splittings motivated by $SU(3)$ chiral symmetry and large $N_c$~\cite{WalkerLoud:2011ab}.  The difficulties with the convergence make the results less convincing than those presented here.}

I would like to return to the connection of $\d M_N$ with BBN.
We now know with some confidence, the two contributions to the mass splitting from first principles
\begin{eqnarray}
M_n - M_p[\textrm{ MeV}] &=& \d M^\g_{n-p} + \d M^{m_d - m_u}_{n-p}\, ,
\nonumber\\
	&=& -178(04)(64) \times \alpha_{f.s.}
		+0.95(8)(6) \times (m_d -m_u)[\mu=2\textrm{ GeV}]\, .
\end{eqnarray}
The electromagnetic self-energy is taken from Ref.~\cite{WalkerLoud:2012bg}.%
\footnote{There are now two published LQCD calculations of this quantity also by Blum et.al. and BMWc~\cite{lattice_isospin}.} 
The quark mass contribution is determined from the current LQCD average of $m_d - m_u$~\cite{Aoki:2013ldr} and the lattice average of $\d M^{m_d-m_u}_{n-p}$ presented above.
A more precise determination of $\d M^\g_{n-p}$ results from combining the experimental result with the lattice calculation of $\d M^{m_d-m_u}_{n-p}$.

\begin{figure}
\center
\includegraphics[width=0.42\textwidth]{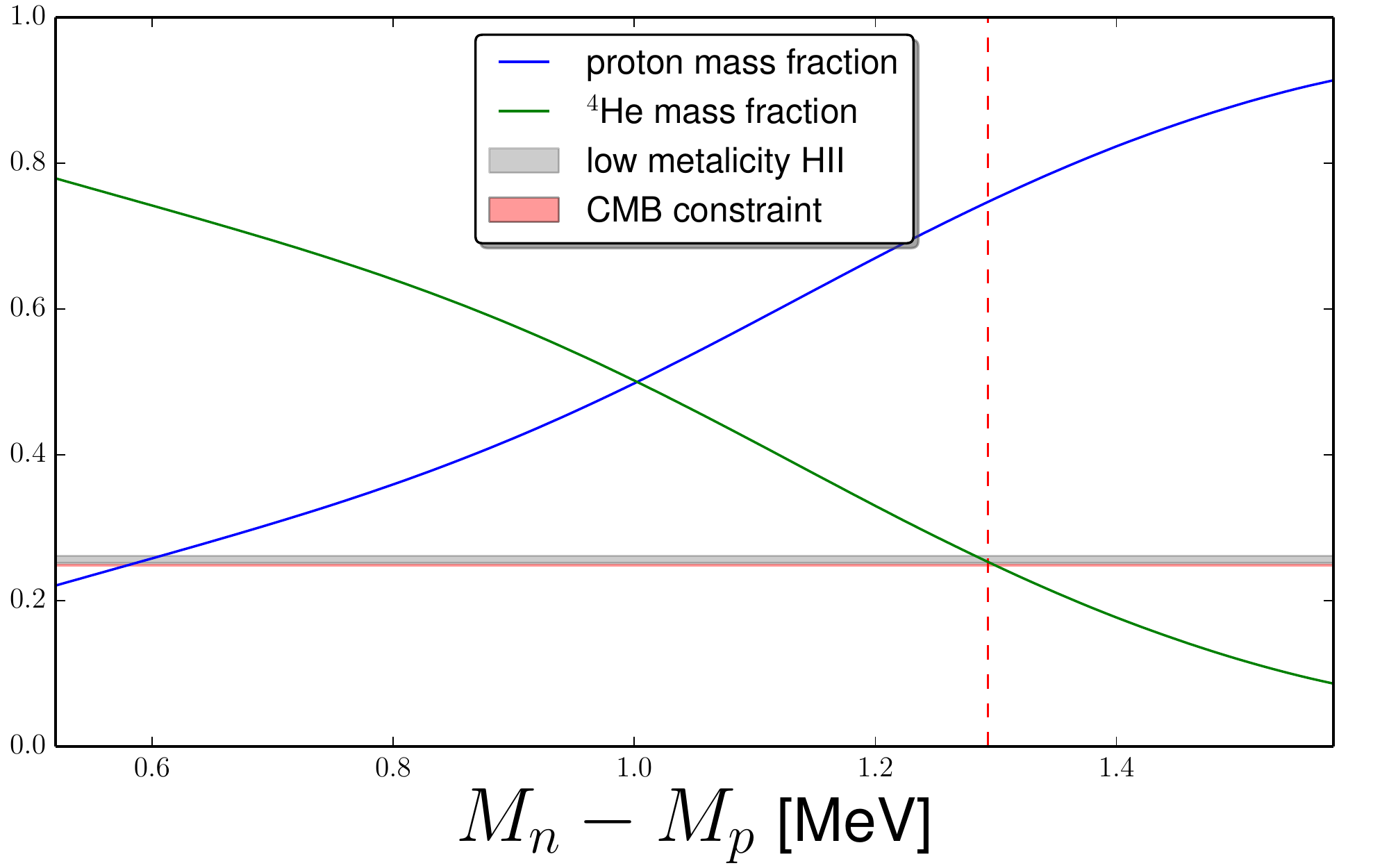}
\includegraphics[width=0.42\textwidth]{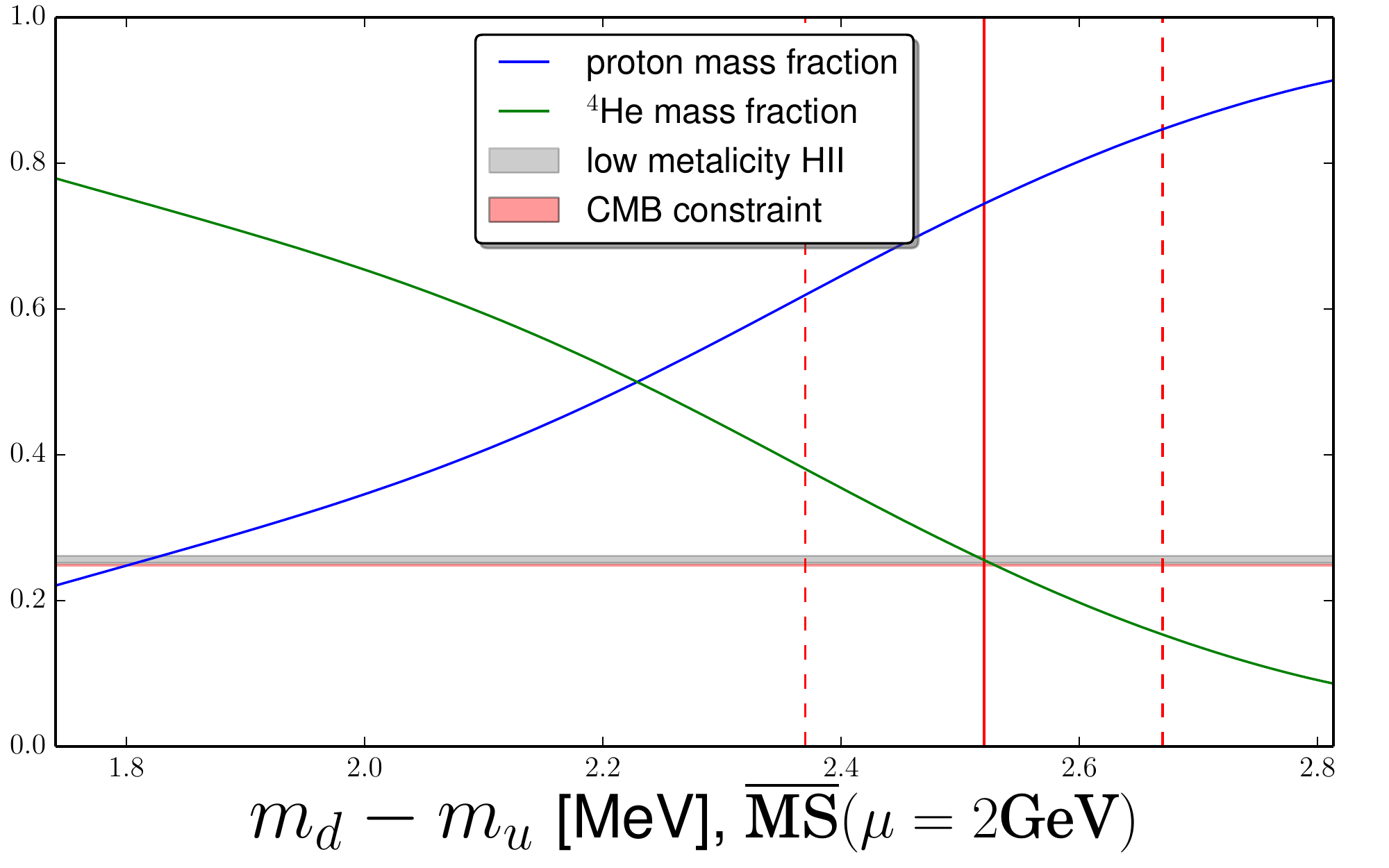}
\caption{\label{fig:bbn_md-mu}Preliminary work relating variation of $m_d - m_u$ to the production of H and ${}^4$He in BBN~\cite{bbn_md-mu}.}
\end{figure}

In addition to the academic interest of making quantitative connections between QCD and the early Universe, we can use BBN to constrain the possible time-variation of fundamental constants.
Considering possible variation of both sources of isospin violation will relax the constraints since they drive the nucleon mass splitting in opposite directions.
But for now, we will freeze the electromagnetic coupling, and consider only the effect of varying the quark mass splitting~\cite{bbn_md-mu}.
We consider only LO isospin breaking corrections so we can ignore variation of the deuteron binding energy.
In Fig.~\ref{fig:bbn_md-mu}, I display the change in the abundances of H and ${}^4$He that result from varying $M_n - M_p$ (left).  Using LQCD, we can relate the x-axis to a change in $m_d - m_u$.
Considering only the uncertainty on $m_d - m_u$ we see the ${}^4$He mass fraction would vary from $\sim$40\% to 10\%, well outside the observed abundance of $\sim$25(1)\%.
It is interesting to note that a precise calculation of how $M_n - M_p$ varies with $m_d-m_u$ could be used to place a tighter constraint on $m_d - m_u$ than presently exists.
This is  a simple example of how LQCD can now be used in nuclear physics to make interesting quantitative connections between the quarks and the cosmos.

\section{Nuclear Physics Review}
I review the current status of LQCD calculations of multi-hadron systems with an emphasis on nuclei.
I refer the reader to the above section as an example of interesting motivation.

\subsection{Methods and Results}
Using LQCD, we are not able to directly compute scattering amplitudes as the calculations are performed in a finite Euclidean volume. 
Moreover, the large Euclidean time behavior of the Green's function is not related to the physical scattering amplitude of interest~\cite{Maiani:1990ca}.
However, it is well known that the infinite volume scattering phase shift can be determined from the dependence of the finite volume energy levels on the spatial volume, a technique developed for interacting quantum field theories by L\"{u}scher; for two particles below inelastic threshold, there is a one-to-one correspondence between the finite volume energy levels and the infinite volume scattering phase shift at the corresponding energy for any unitary theory, up to corrections which vanish exponentially in the volume~\cite{Luscher:1986pf}, e.g.~\cite{Bedaque:2006yi}.

\subsubsection{L\"{u}scher Method}
To determine the scattering amplitude, one first computes the energy levels of the one and two particle states which allows for a determination of the interacting momentum
\begin{equation}
E = 2\sqrt{m^2 + k^2}\, .
\end{equation}
In the absence of interactions, $k \in 2\pi \vec{n}/L$ with $\vec{n} \in \mathbb{Z}^3$.
One then solves for the phase shift%
\footnote{The right hand side of the equation is a representation of the Riemann-Zeta function, which is valid for $S$-wave scattering in a $3d$ spatial volume of size $L$ with periodic boundary conditions, in the $A_1$ representation, ignoring the partial-wave mixing induced by the cubic box.
There has been a significant amount of formal development to extend our understanding to include boosted systems, higher partial waves and coupled channels~\cite{fv_formalism}.
}
\begin{equation}
k\cot \delta(k) = 
	\frac{1}{\pi L} \left( \sum_{|\vec{n}| < \Lambda} 
	\frac{1}{\vec{n}^2 -\frac{k^2 L^2}{4\pi^2}} - 4\pi \Lambda
	\right)\, .
\end{equation}
\begin{figure}
\center
\includegraphics[width=0.4\textwidth]{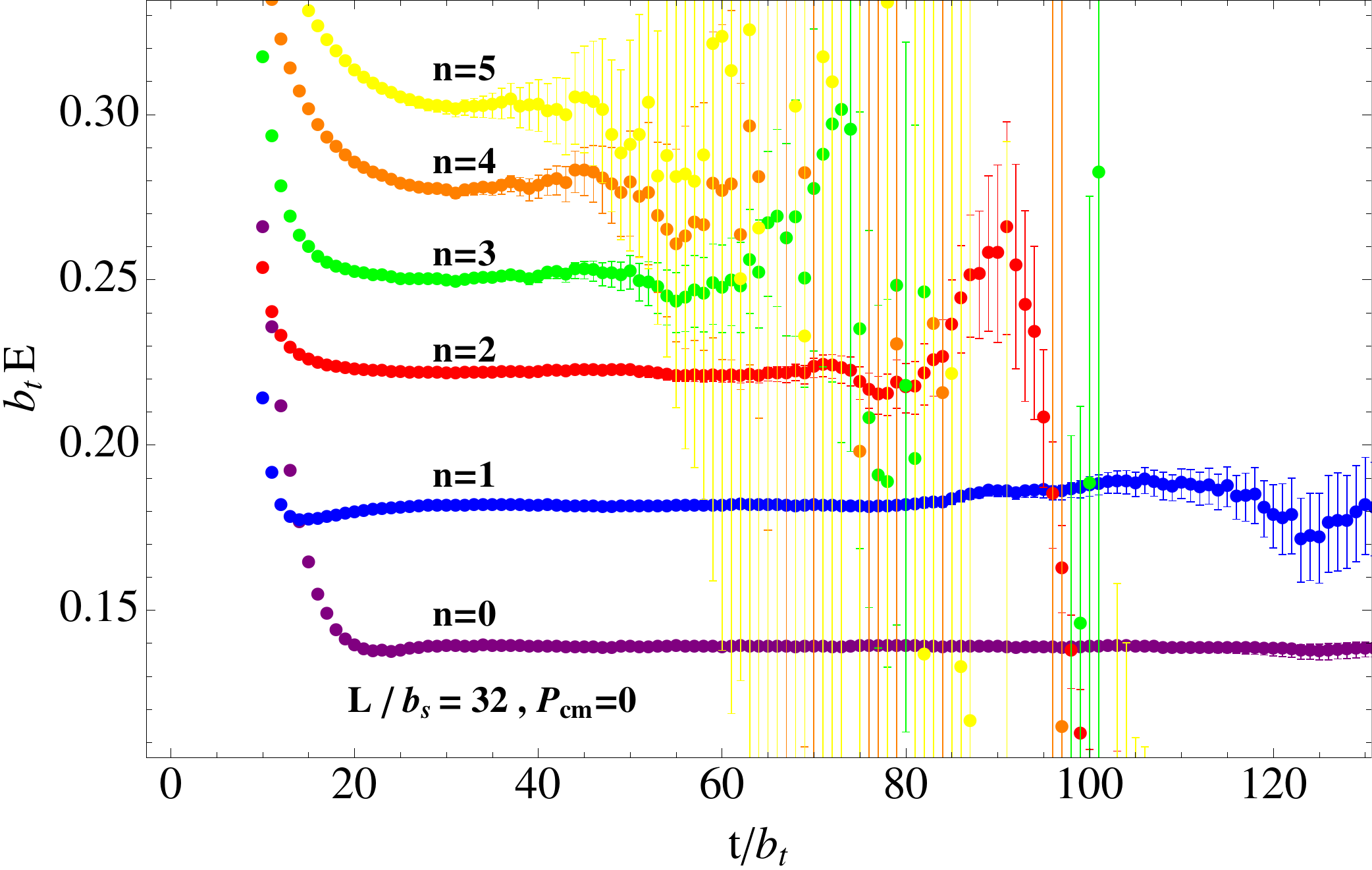}
\includegraphics[width=0.4\textwidth]{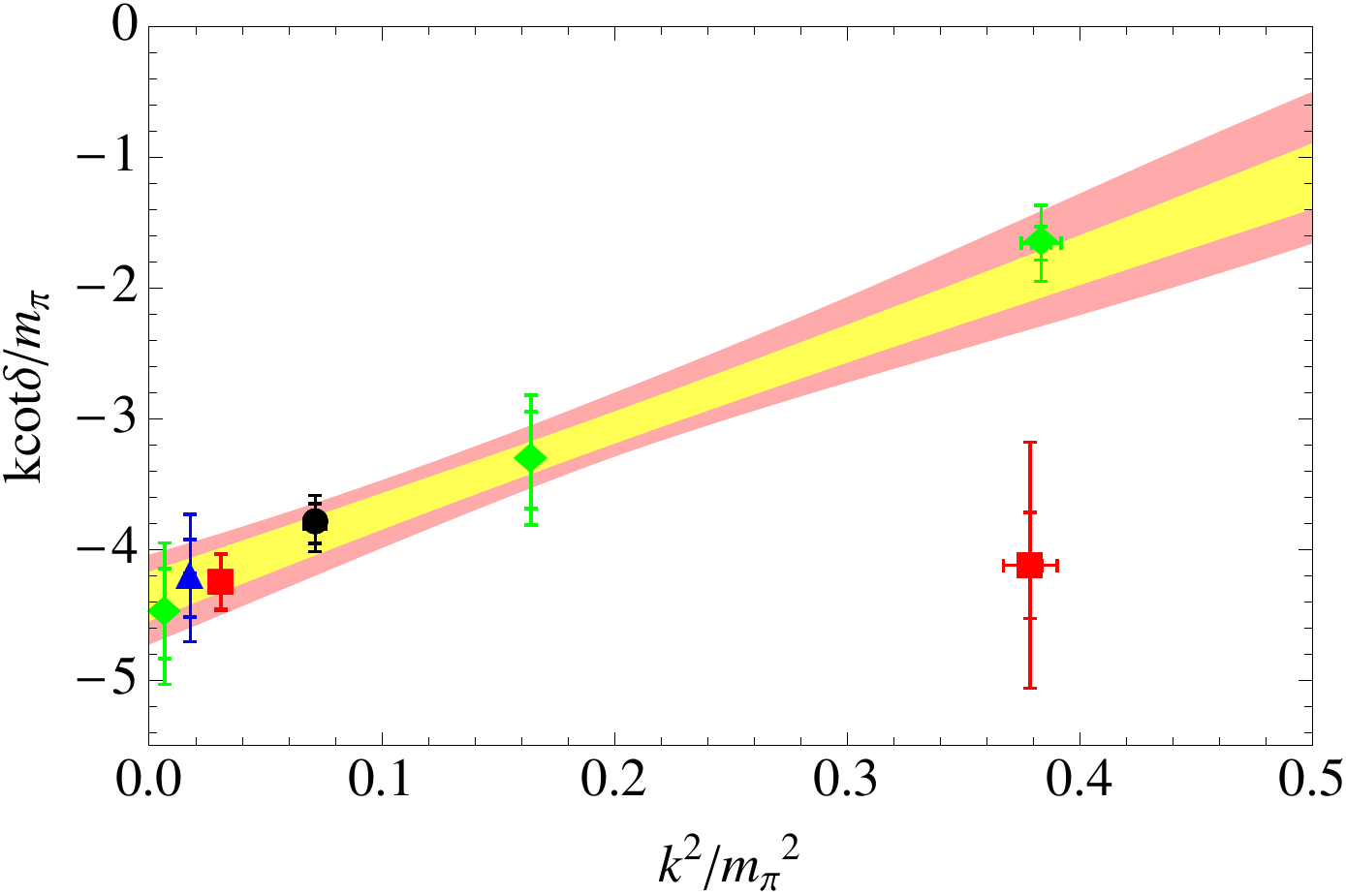}
\caption{\label{fig:nplqcd_I2_pipi}Calculation of the $I=2$ $\pi\pi$ phase shift by NPLQCD at $m_\pi \sim 400$~MeV~\cite{Beane:2011sc}.}
\end{figure}

With several energy levels, one can then parameterize the phase shift with the effective range expansion, valid for small $k$,
\begin{equation}
k \cot \delta(k) = -\frac{1}{a} + \frac{1}{2} r k^2 +\dots
\end{equation}
where $a$ and $r$ are the scattering length and effective range respectively.  As an example, in Fig.~\ref{fig:nplqcd_I2_pipi}, I present a recent calculation by NPLQCD of the $I=2$ $\pi\pi$ phase shift with $m_\pi\sim400$~MeV~\cite{Beane:2011sc}.
For a calculation of the same quantity, on the same gauge configurations, with a more sophisticated set of operators and the full variational method, see Ref.~\cite{Dudek:2012gj}.

\subsubsection{HAL QCD Method 1: Bethe-Salpeter wave-function \label{sec:halqcd_1}}
\begin{figure}
\center
\begin{tabular}{ccc}
\includegraphics[width=0.3\textwidth]{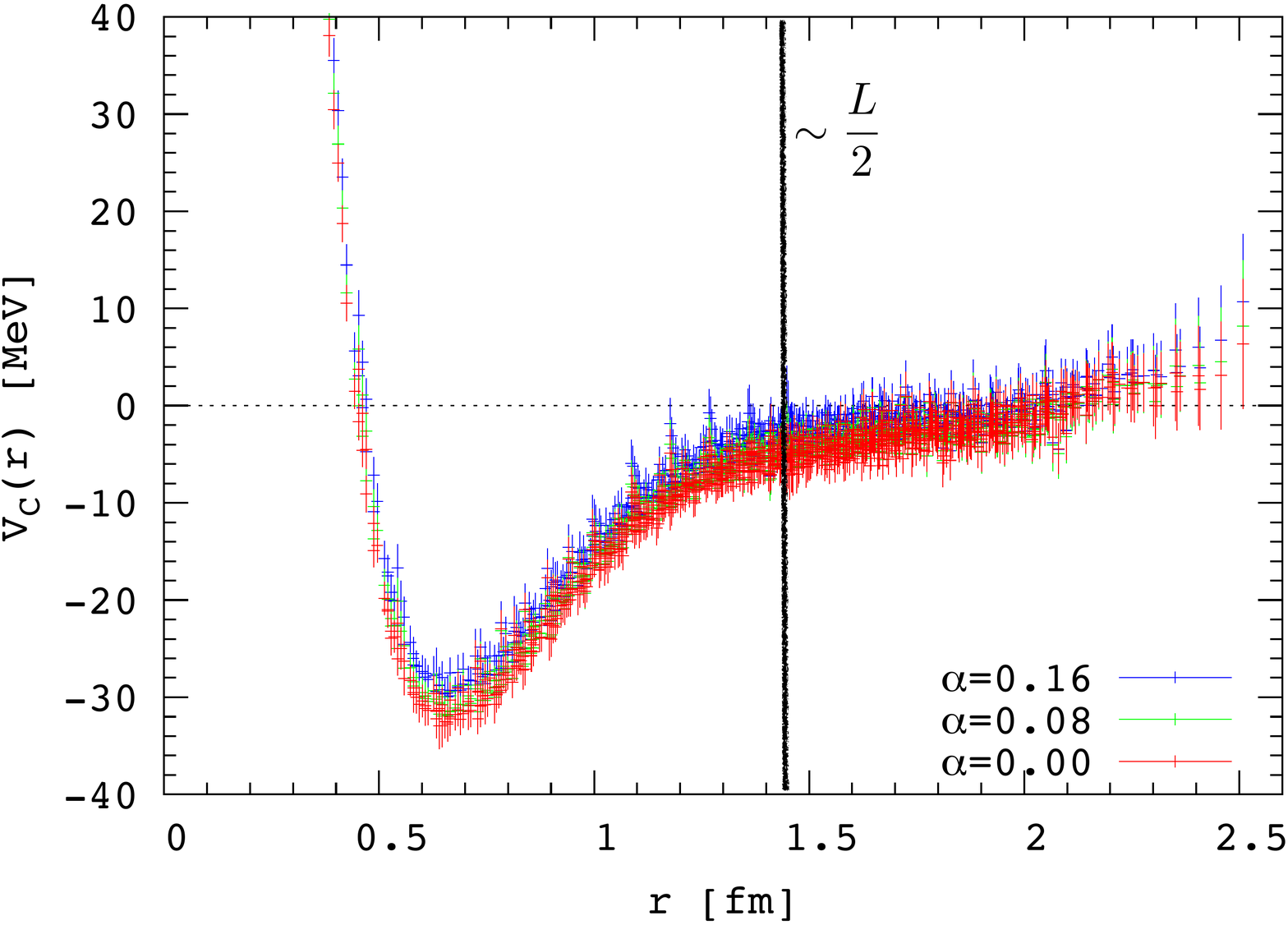}
&\includegraphics[width=0.31\textwidth]{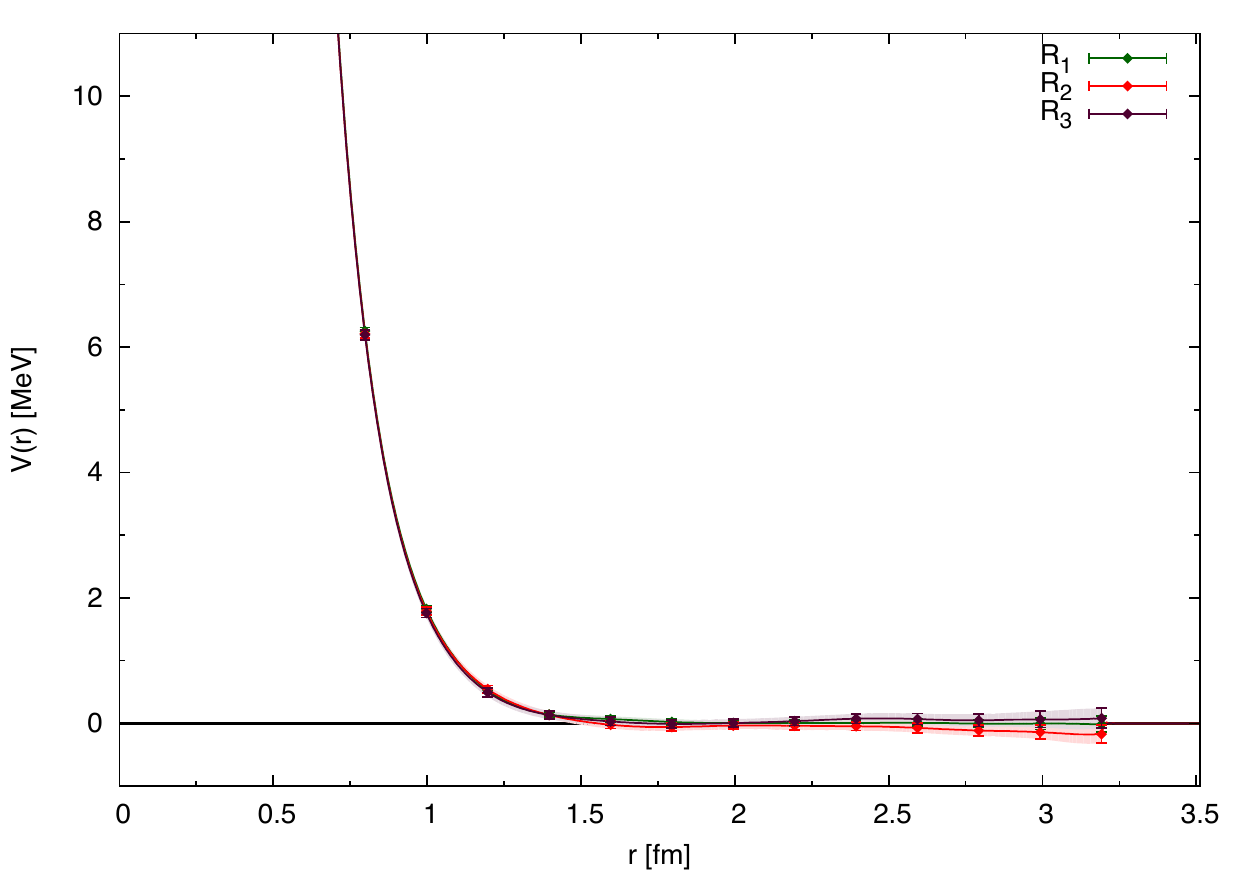}
&\includegraphics[width=0.33\textwidth]{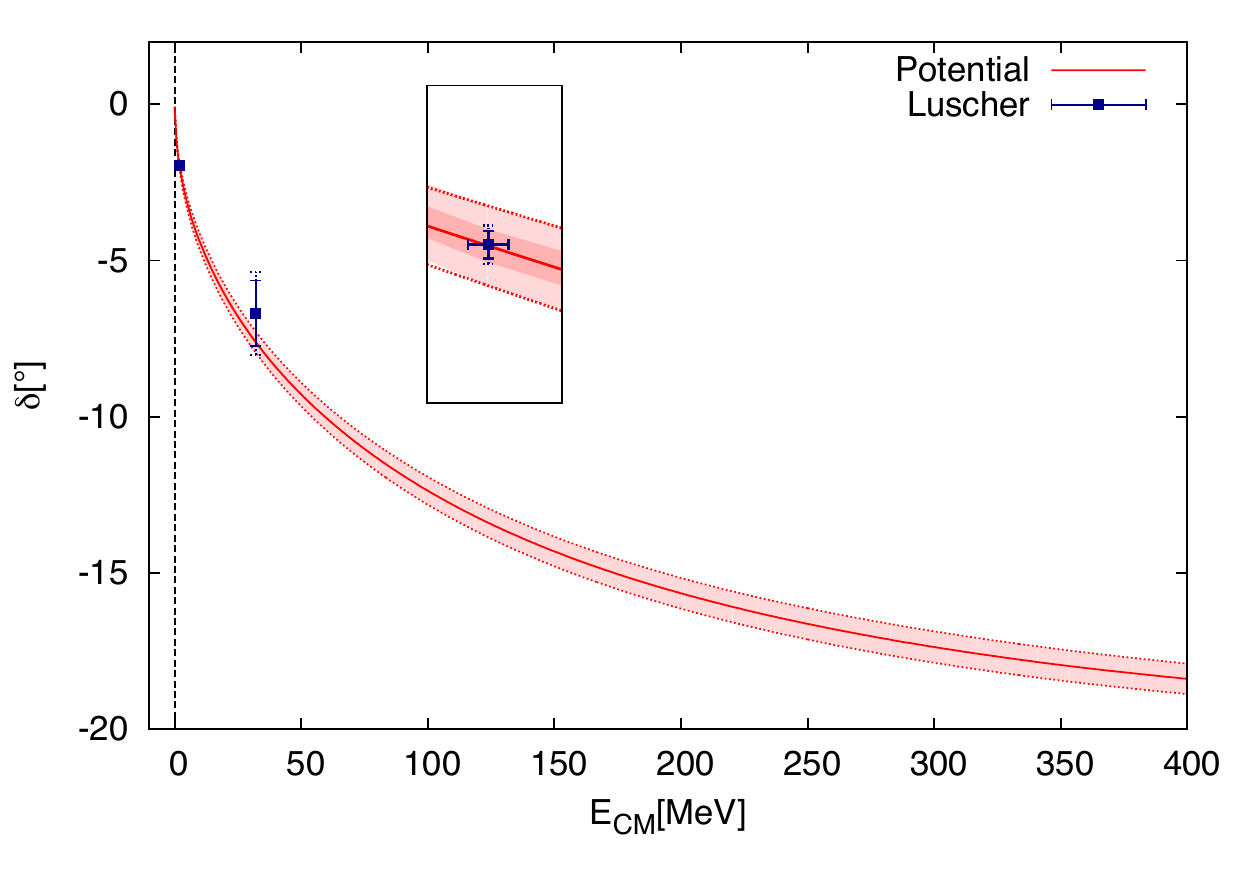}
\\
$NN$ potential & $I=2\ \pi\pi$ potential & $I=2\ \pi\pi$ phase shift
\end{tabular}
\caption{\label{fig:halqcd_pot}Sample $NN$ potential computed by HALQCD taken from Ref.~\cite{HALQCD:2012aa} (left).  $I=2\ \pi\pi$ potential using the time-dependent potential method (middle) which is used to determine the phase shift and compared with the L\"{u}scher method (right) from Ref.~\cite{Kurth:2013tua}.}
\end{figure}

Recently, an alternative method has been advocated by the HALQCD Collaboration~\cite{Ishii:2006ec} which utilizes the Bethe-Salpeter wave-function, as first applied to the $I=2\ \pi\pi$ system~\cite{Aoki:2005uf};
\begin{equation}
	\left[ \frac{\mbf{p}^2}{2\mu} - H_0 \right] \psi_\mbf{p}(\mbf{r})
	=
	\int d^3 r^\prime U(\mbf{r},\mbf{r}^\prime) \psi_\mbf{p}(\mbf{r^\prime})
\end{equation}
$H_0 = -\nabla^2 / 2\mu$ and in the absence of interactions $H_0 \psi_\mbf{p}(\mbf{r}) = \frac{\mbf{p}^2}{2\mu} \psi_\mbf{p}(\mbf{r})$.
A choice for the finite volume Bethe-Salpeter wave-funciton is 
\begin{equation}\label{eq:BS}
\psi_\mbf{p}(\mbf{r}) =
	\frac{1}{V} \sum_{\mbf{x}} \langle 0 \Big| N(\mbf{x} +\frac{\mbf{r}}{2}) N(\mbf{x} -\frac{\mbf{r}}{2})
		\Big| N(\mbf{p}) N(-\mbf{p}) \rangle_{in}\, ,
\end{equation}
where $|N(\mbf{p}) N(-\mbf{p}) \rangle_{in}$ is an incoming two-nucleon state with center-of-mass momentum $|\mbf{p}|$.
Consider the two-particle correlation function
\begin{eqnarray}
C_{NN}(\mbf{r},t) &=& \sum_\mbf{x} \langle 0 \Big| 
		N(\mbf{x} +\frac{\mbf{r}}{2},t) N(\mbf{x} -\frac{\mbf{r}}{2},t)
		N^\dagger(\mbf{x}_0,0) N^\dagger(\mbf{x}_0,0)
		\Big| 0 \rangle
\nonumber\\
	&=& \sum_n \sum_\mbf{x} e^{-E_n t} \langle 0 \Big|
		N(\mbf{x} +\frac{\mbf{r}}{2},0) N(\mbf{x} -\frac{\mbf{r}}{2},0) \Big| n \rangle
		\langle n \Big| N^\dagger(\mbf{x}_0,0) N^\dagger(\mbf{x}_0,0)
		\Big| 0 \rangle
\nonumber\\
	&=& \sum_n e^{-E_n t} \psi_n(\mathbf{r}) A^\dagger_n\, .
\end{eqnarray}
In the long time limit, the Bethe-Salpeter wave function of interest \eqref{eq:BS} is recovered.
A simple sum recovers the correlation function used in the standard L\"{u}scher method for total momentum $\mathbf{P}$
\begin{equation}
	C_{NN}(\mbf{P},t) = \sum_\mbf{r} e^{i\mbf{P}\cdot\mbf{r}} C_{NN}(\mbf{r},t)\, .
\end{equation}
The next step taken with this method is to approximate the Bethe-Salpeter potential with a local potential and a gradient expansion
\begin{equation}
\label{eq:local_pot}
U(\mbf{r},\mbf{r}^\prime) =
	V_C(\mbf{r}) \delta(\mbf{r}-\mbf{r}^\prime) + \mc{O}(\nabla_\mbf{r}^2/\L^2)\, ,
\end{equation}
where $\L$ is a bit ambiguous.
This approximated potential can then be numerically computed with the correlation function designed to isolate the Bethe-Salpeter wave-function,
\begin{eqnarray}
\label{eq:halqcd_pot}
V_C(\mbf{r}) &\simeq& \frac{\mbf{p}^2}{2\mu} 
	+\lim_{t\rightarrow \infty} \frac{1}{2\mu} \frac{\nabla_\mbf{r}^2 C_{NN}(\mbf{r},t)}{C_{NN}(\mbf{r},t)}
	= \frac{\mbf{p}^2}{2\mu}
	+\frac{1}{2\mu} 
	\frac{\nabla_\mbf{r}^2 (e^{-E_0 t} \psi(\mbf{r}) A^\dagger_0)}{e^{-E_0 t} \psi(\mbf{r}) A^\dagger_0}
\nonumber\\
	&\simeq& 
	\frac{\mbf{p}^2}{2\mu}
	+\frac{1}{2\mu} \frac{\nabla_\mbf{r}^2 \psi(\mbf{r})}{\psi(\mbf{r})}\, .
\end{eqnarray}
There are several assumptions, approximations and challenges with this method:
\begin{itemize}
\item The resulting uncertainty from the approximation to the potential~\eqref{eq:local_pot} is difficult to quantify, as the $\nabla_\mbf{r}^2 / \L^2$ expansion is not systematically improvable in the same sense as an EFT.

\item The periodic images must be accounted for in determining potential (HALQCD does include image potentials in their analysis).  See Fig.~\ref{fig:halqcd_pot} (left) for a sample potential computed by HALQCD~\cite{HALQCD:2012aa}.  I have added a vertical line to indicate the location of $L/2$.

\item It is misleading to plot the potential for $r< 1/\D E^*$ as this region is polluted by inelastic contributions, yet the figure (left of Fig.~\ref{fig:halqcd_pot}) leads to ``warm and fuzzy'' feelings as we all ``know'' this is what nuclear potentials look like (tongue in cheek).
\end{itemize}

\subsubsection{HAL QCD Method 2: time-dependent Schr\"{o}dinger-like equation \label{sec:time_dep_pot}}
The HALQCD potential method is susceptible to the same long-time stochastic noise problems that plague all lattice calculations of multi-baryon systems.
They have therefore developed a new ``time-dependent'' Schr\"{o}dinger-like equation with the aim of extracting information from the correlation function earlier in Euclidean time~\cite{HALQCD:2012aa}.  
\begin{equation}
\left[ \frac{1}{4M} \partial_t^2 -\partial_t -H_0 \right] R(\mbf{r},t)
	= \int d^3 r^\prime U(\mbf{r},\mbf{r}^\prime) R(\mbf{r}^\prime, t),
	\qquad
	R(\mbf{r},t) = \frac{C_{NN}(\mbf{r},t)}{(C_N(t))^2}
\end{equation}
The key observation is that if the correlation function only contains support from elastic states, the same non-local potential $U(\mbf{r},\mbf{r}^\prime)$ describes all the scattering states, not just the ground state.
The strategy is to take $t$ sufficiently large that only the ground state contributes to $C_N(t)$.
One then applies the same gradient approximation for the potential \eqref{eq:local_pot}.  The resulting local potential can then be determined with the equation
\begin{equation}
\label{eq:time_dep_pot}
V_C(\mbf{r}) \simeq
	\frac{1}{M} \frac{\nabla_\mbf{r}^2 R(\mbf{r},t)}{R(\mbf{r},t)}
	-\frac{\partial_t R(\mbf{r},t)}{R(\mbf{r},t)}
	+\frac{1}{4M} \frac{\partial_t^2 R(\mbf{r},t)}{R(\mbf{r},t)}\, .
\end{equation}
Recently, HALQCD is making progress in comparing their new method with the standard L\"{u}scher method.
In Ref.~\cite{Kurth:2013tua}, a detailed comparison was made using the $I=2\ \pi\pi$ system in a quenched calculation with $m_\pi\sim940$~MeV.
This time-dependent method was used to compute the potential, and then solve the infinite volume Schr\"{o}dinger equation, which can be used to determine the phase shift for continuous values of the interacting energy (momentum).
Fig.~\ref{fig:halqcd_pot} displays the $I=2\ \pi\pi$ potential and the resulting phase shift, as well as a comparison with the L\"{u}scher method.
Good agreement was found between both methods.
At this conference, further comparison was performed in the $I=2\ \pi\pi$ system~\cite{Charron:2013paa} using the variational method~\cite{Luscher:1990ck}, also finding good agreement between the methods.%
\footnote{Below the inelastic threshold, the $s$-channel diagrams which give rise to the power-law volume dependence of the $I=2\ \pi\pi$ system, are free of the unitarity-violating effects which arise from quenched and partially-quenched theories, and so one expects the L\"{u}scher relation to hold.  See Ref.~\cite{Chen:2005ab} for further discussion.}

While these comparisons are encouraging, there still remain significant assumptions, approximations and challenges with this method:
\begin{itemize}
\item The same issues with the time-independent potential method also apply here, Sec.~\ref{sec:halqcd_1}

\item The $I=2\ \pi\pi$ system is particularly special and simple (weakly repulsive with a large gap to inelastic states) and is not a good test for the difficulties encountered in the $NN$ system

\item The assumption is that the correlation function is free from contamination from inelastic states.
It is challenging to demonstrate
\begin{equation}
	C_{NN}(\mbf{r},t) = \sum_{n \in \textrm{elastic}} e^{-E_n t} \psi_n(\mbf{r}) A^\dagger_n\, ,
\end{equation} 
but without such a proof, an unquantifiable systematic is introduced.  
For example, contributions from inelastic states would mean the potential $U(\mbf{r},\mbf{r}^\prime)$ as determined from Eq.~\eqref{eq:time_dep_pot} would be polluted in ways not necessarily parameterized by the gradient expansion of Eq.~\eqref{eq:local_pot}.
\end{itemize}

\subsubsection{Results}
Lattice QCD calculations of multi-baryon systems began in earnest in 2006 with the first dynamical lattice calculation of the $NN$ system by NPLQCD~\cite{Beane:2006mx}, followed by the quenched calculation by Ishii et. al. (who would become HALQCD)~\cite{Ishii:2006ec}.
These calculations occurred a little over a decade after the pioneering quenched calculations of Fukujita et. al.~\cite{Fukugita:1994ve}.
The first three (and higher) body calculations began in 2008 for mesons~\cite{Beane:2007es} and 2009 for baryons~\cite{Beane:2009gs}.
In 2010, Yamazaki et. al. (members of PACS-CS) joined the effort by first demonstrating the number of Wick contractions for ${}^{\{3,4\}}$He could be reduced from $\{2880,\, 518400\} = \{5!\times4!,\, 6!\times6!\}$ to \{93,\, 1107\} respectively, by taking advantage of all the symmetries in these systems.  The new contraction algorithm was used to demonstrate the existence of bound ${}^{\{3,4\}}$He nuclei in quenched lattice calculations with $m_\pi \sim 800$~MeV~\cite{Yamazaki:2009ua}.
In the last couple years, there has been a significant growth in the number of people/groups thinking about LQCD calculations of multi-nucleon (baryon) systems.
This is correlated with the significant growth in available computing resources dedicated to this area of research and the general realization that we (the lattice community) have a real opportunity to make significant contributions to our understanding of nuclear physics with LQCD.

The end of 2010 also marked a significant milestone for LQCD calculations of multi-baryon systems.
the first dynamical LQCD calculations of a bound multi-baryon system was performed:
both NPLQCD~\cite{Beane:2010hg} and HALQCD~\cite{Inoue:2010es} performed calculations demonstrating the existence of a bound H-dibaryon with heavier than physical pion masses.
The H-dibaryon was first proposed by R.~L.~Jaffe~\cite{Jaffe:1976yi} as a strongly attractive channel, having the quantum numbers of an $SU(3)$ flavor singlet in the strangeness -2 sector,
\begin{equation}
|H\rangle = 
	-\sqrt{\frac{1}{8}} | \L \L \rangle + \sqrt{\frac{3}{8}} | \S \S \rangle + \sqrt{\frac{4}{8}} | N \Xi \rangle\, .
\end{equation}
To date, there is no experimental confirmation of a bound h-dibaryon but there hints of interesting behavior and it is an active area of experimental investigation~\cite{h:experiment}.
Nevertheless, this was exciting as it signaled the beginning of the era of LQCD calculations of bound multi-baryon systems.

\begin{figure}
\center
\includegraphics[width=0.35\textwidth]{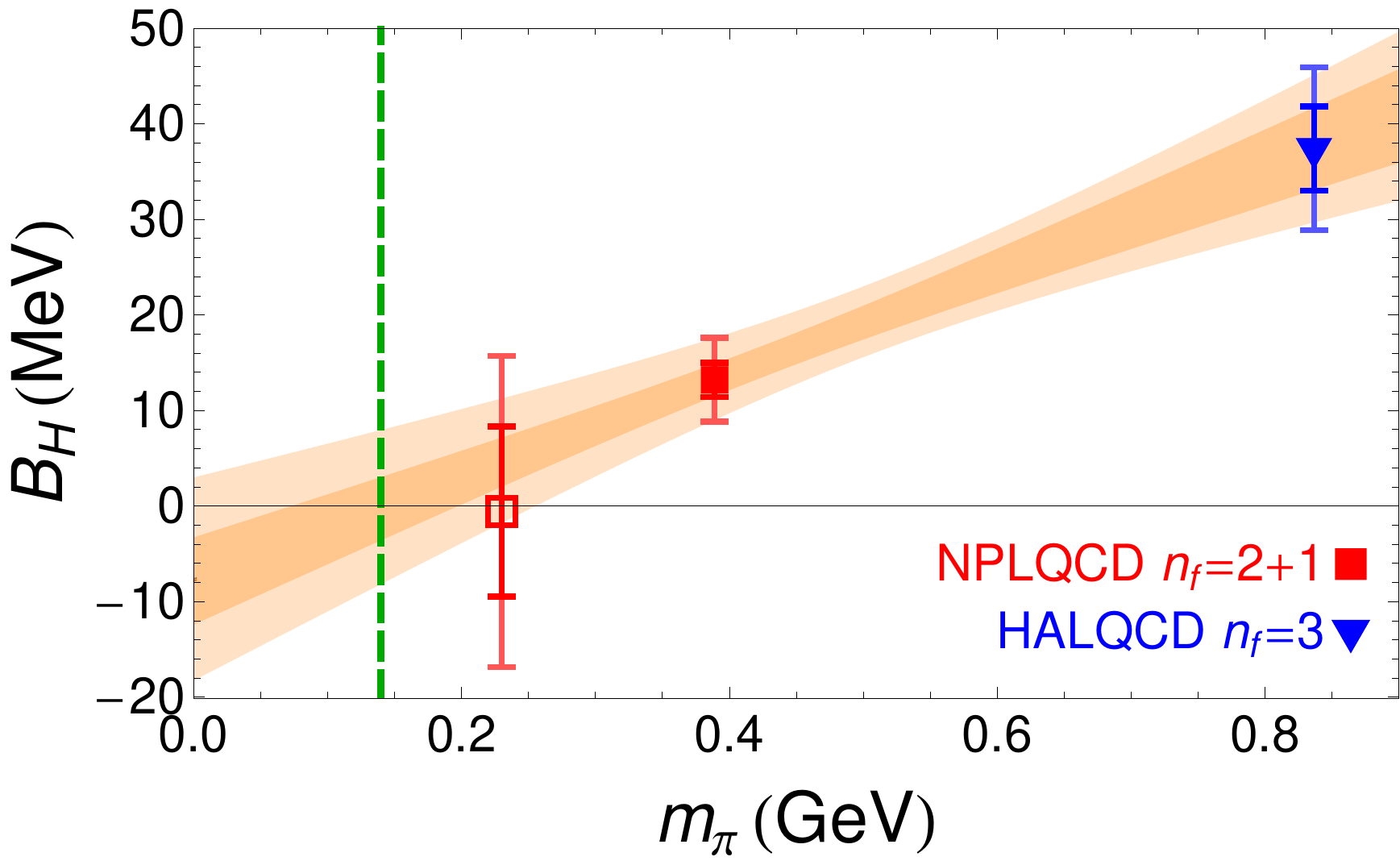}
\includegraphics[width=0.35\textwidth]{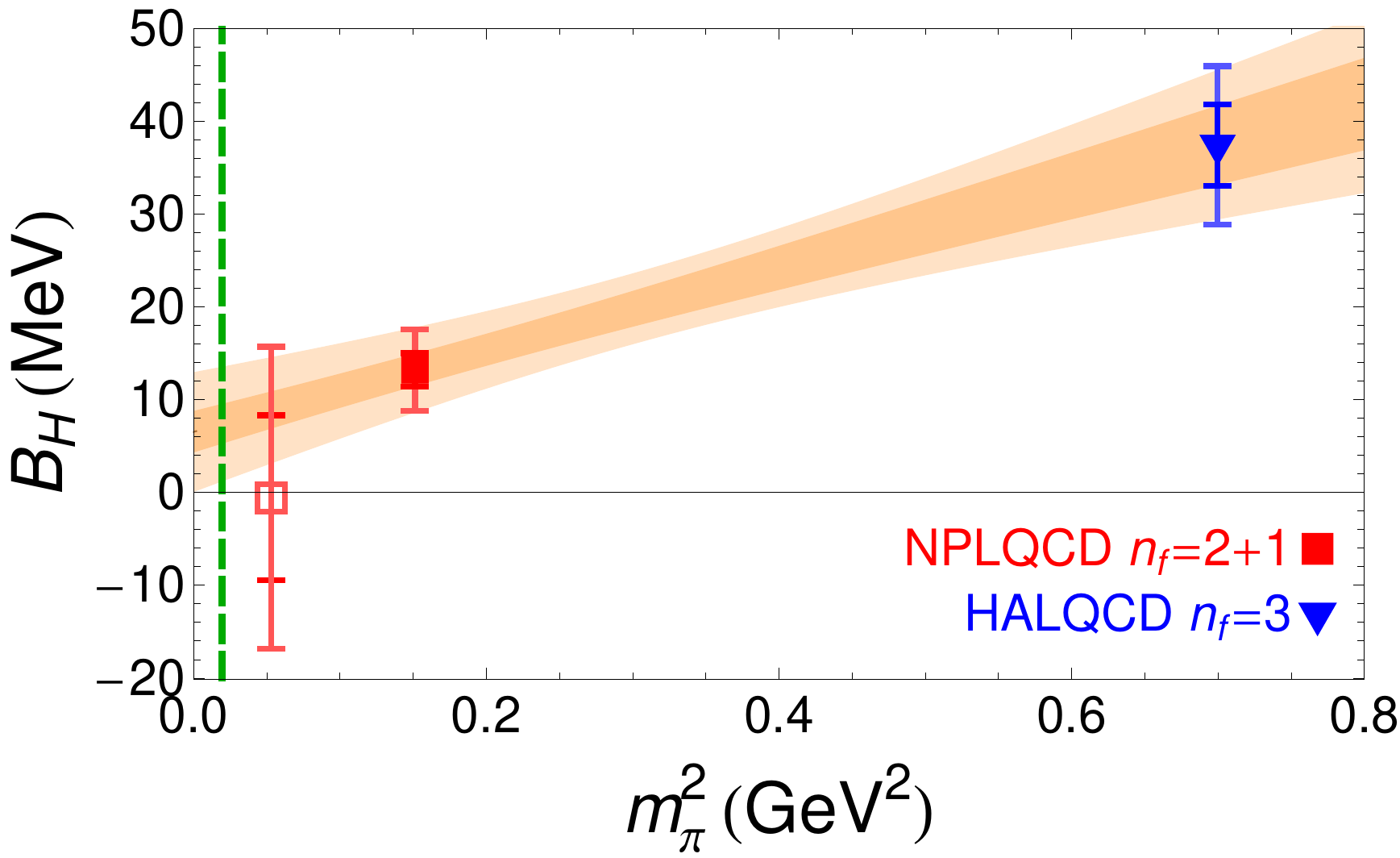}
\caption{\label{fig:h_dibaryon}Estimate of the h-dibaryon binding energy dependence on the pion mass~\cite{Beane:2011zpa}.}
\end{figure}

Given the significant challenge of reducing the pion mass in these calculations, it is worth comparing the results of NPLQCD and HALQCD to get a sense of pion mass dependence of the binding energy.
Fig.~\ref{fig:h_dibaryon} displays two crude estimates of the pion mass dependence of $B_H$ from \cite{Beane:2011zpa}.
The vertical dashed (green) line is at the physical pion mass.  
It is interesting/amusing to note the linear pion mass dependence is consistent with an EFT estimate~\cite{Haidenbauer:2011ah}, perhaps mimicking the linear pion mass dependence of $M_\L$~\cite{WalkerLoud:2008bp}.
As can be seen, the extrapolated binding energy is consistent with zero at the physical pion mass, which is consistent with experimental results.
Loosely bound states (small binding energy) will be more susceptible to the finite size of the box as their wave-functions will exponentially spread out as $\psi(r) \propto e^{-\sqrt{B M} r}$.  
Thus, resolving the ground state energy in LQCD calculations of the h-dibaryon and deuteron for example, will be significantly more challenging than for more deeply bound nuclei, such as ${}^{\{3,4\}}$He.
At this conference, we see an additional group is seriously exploring the h-dibaryon~\cite{Francis:2013lva}.

\subsubsection*{$NN$ systems}
It has long been known that the low-energy $NN$ interactions are finely tuned in both the ${}^1 S_0$ and ${}^3 S_1$(deuteron) channels, as evidenced by their large scattering lengths
\begin{equation}
a_{{}^1 S_0} \simeq -24 \textrm{ fm}\, , \qquad
a_{{}^3 S_1} \simeq 5.5 \textrm{ fm}\, , \qquad
R_{NN} \simeq 1/m_\pi \simeq 1.4 \textrm{ fm}\, .
\end{equation}
Phenomenologically, this is understood to arise from a delicate cancellation between the long, medium and short range nuclear interactions.
These are just two of many examples of fine tunings observed in low-energy nuclear physics, which have a big impact on the Universe we live in, as discussed in Sec.~\ref{sec:KW}.
It will be very interesting when lattice calculations can resolve the nature of this fine tuning and we can in turn propagate that information to our understanding of the early Universe.
This will require high precision calculations to be performed with $m_\pi \lesssim 300$~MeV which will allow for contact with the low-energy $NN$ EFT.

\begin{figure}
\center
\includegraphics[width=0.32\textwidth]{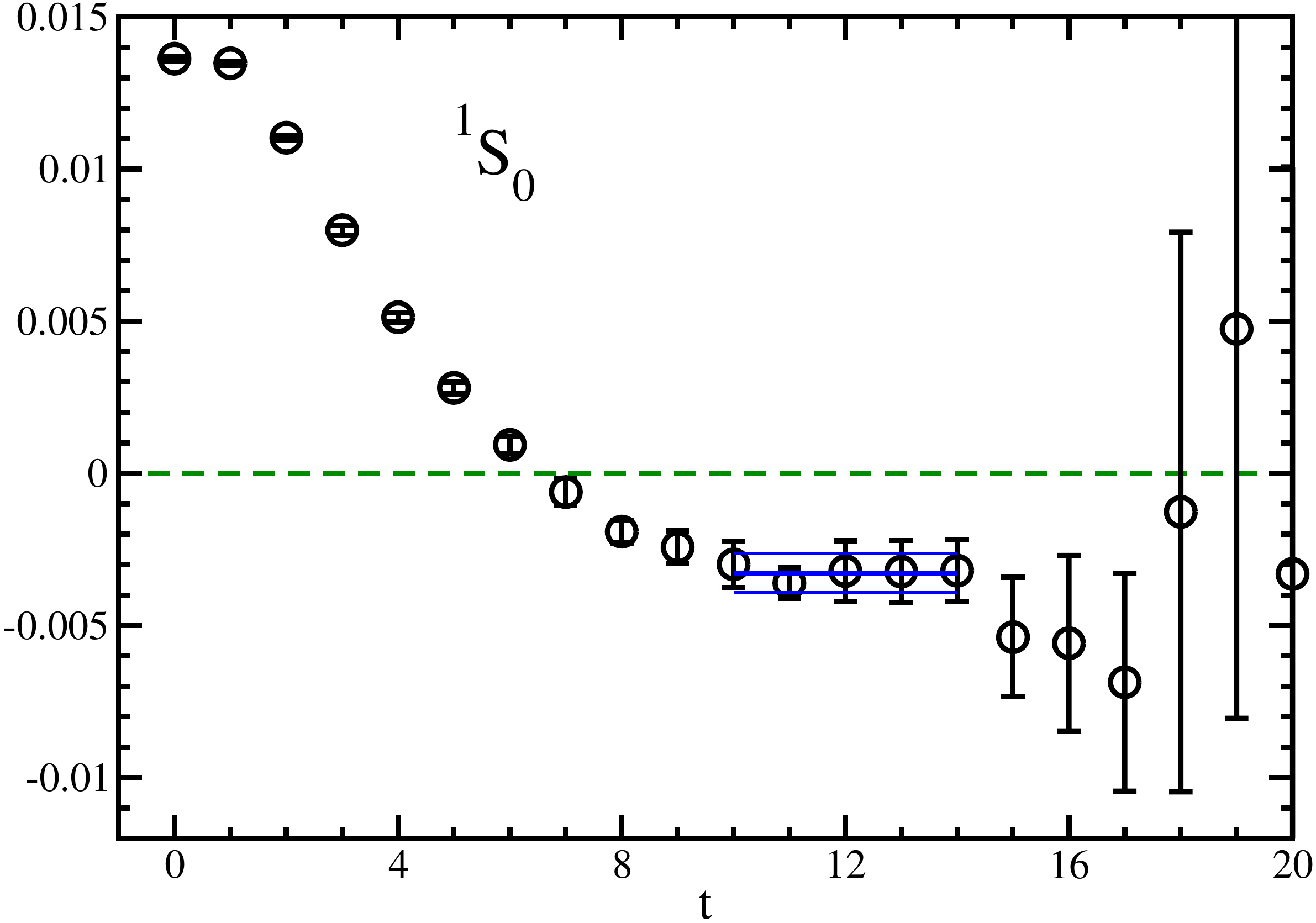}
\includegraphics[width=0.35\textwidth]{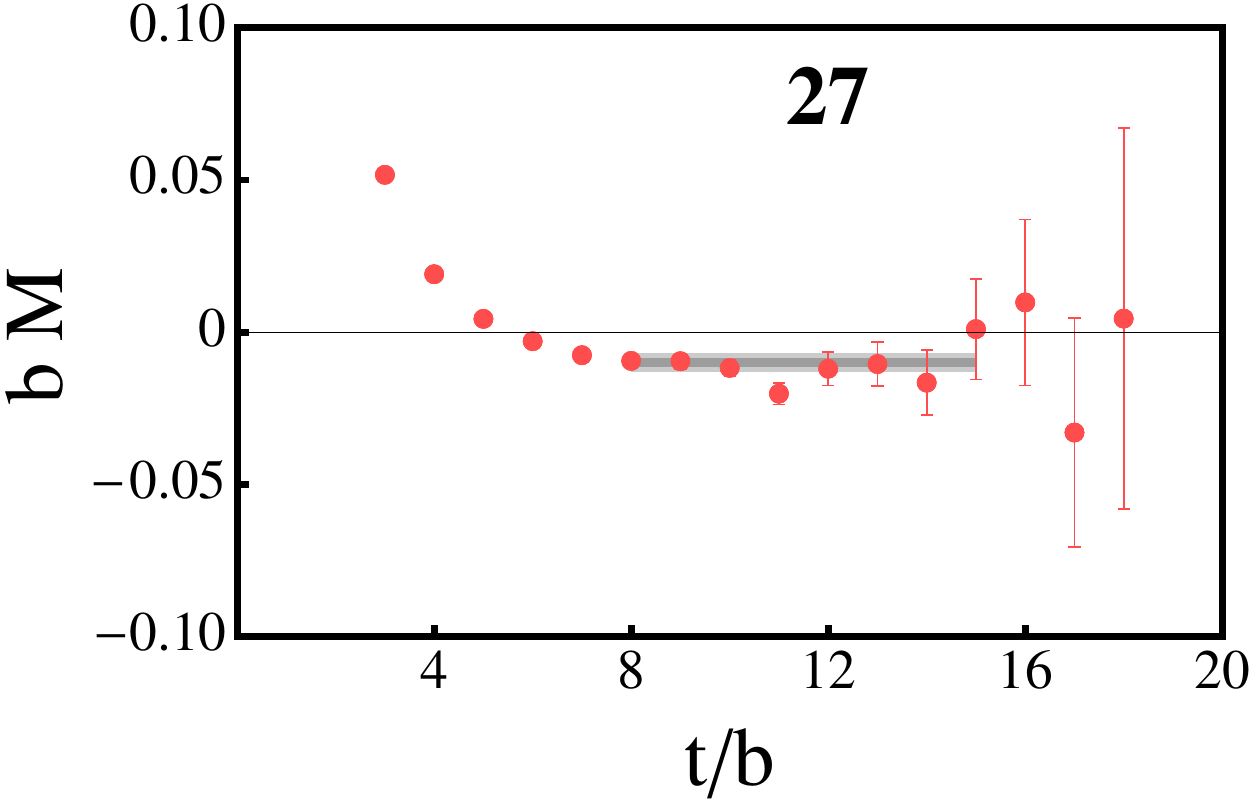}
\caption{\label{fig:nn_bound}Sample effective mass plots of the di-neutron from \cite{Yamazaki:2012hi} (left) and \cite{Beane:2012vq} (right) respectively.}
\end{figure}

\begin{figure}
\center
\begin{tabular}{ccc}
\includegraphics[width=0.37\textwidth]{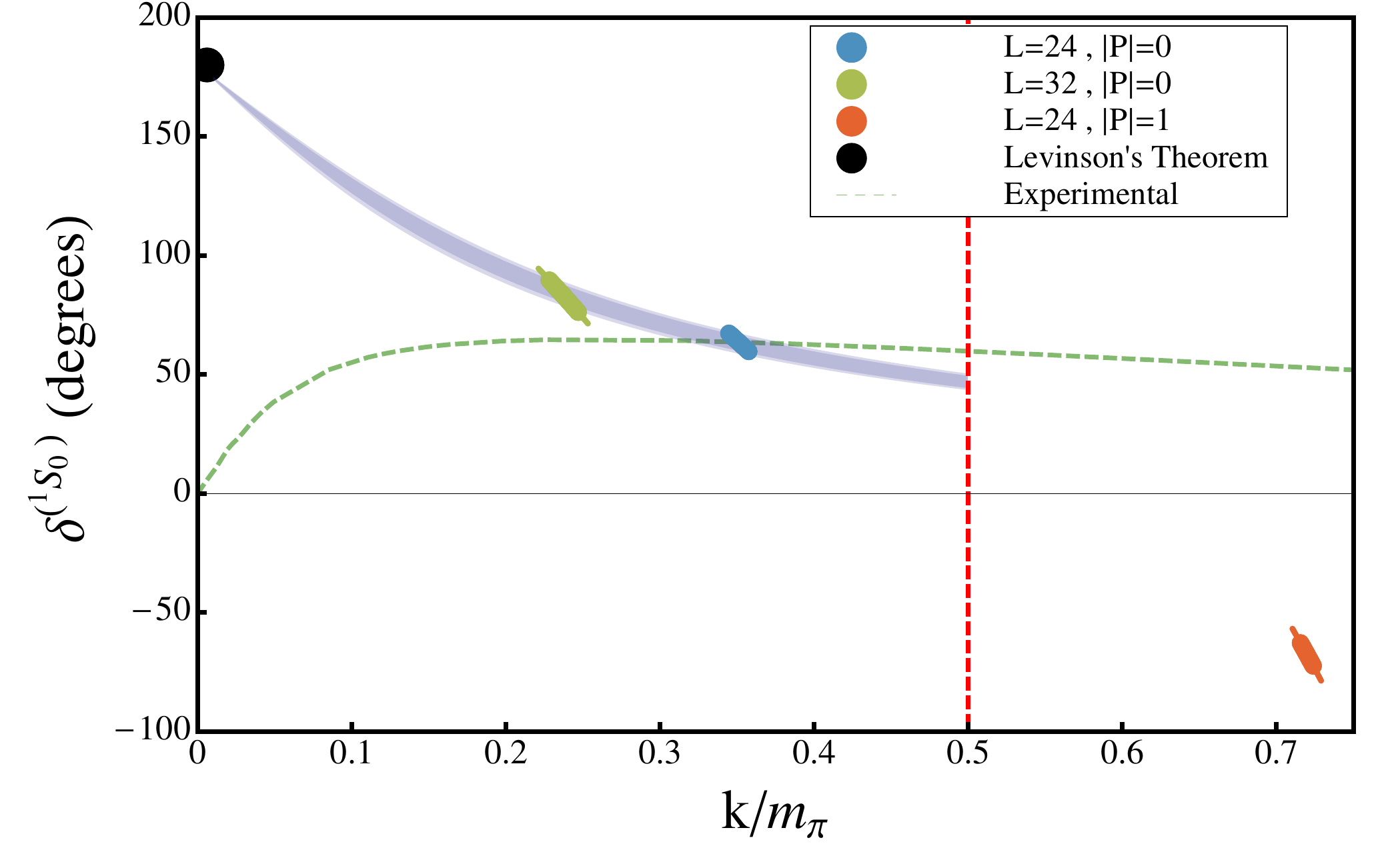}
&\includegraphics[width=0.22\textwidth]{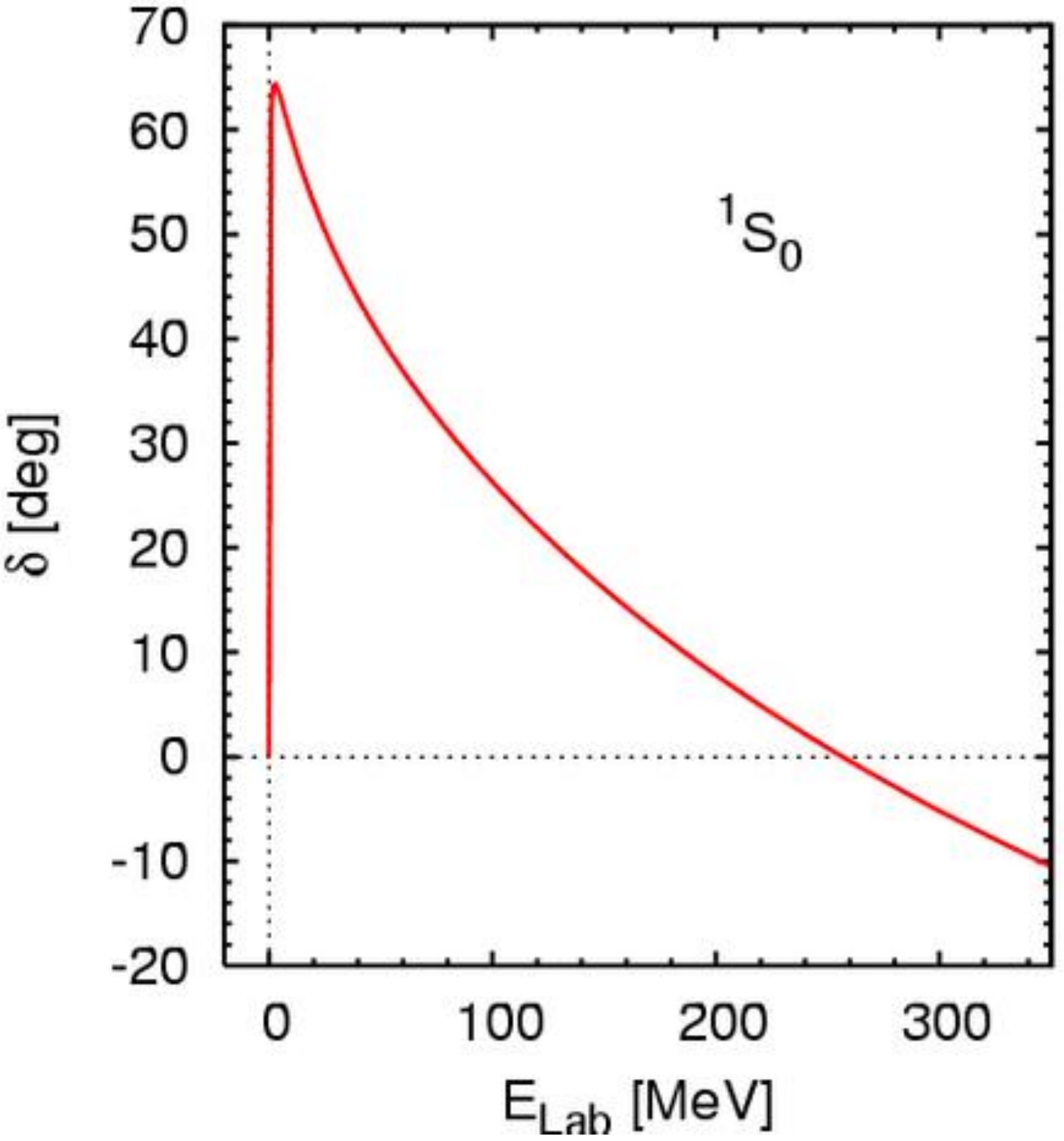}
&\includegraphics[width=0.35\textwidth]{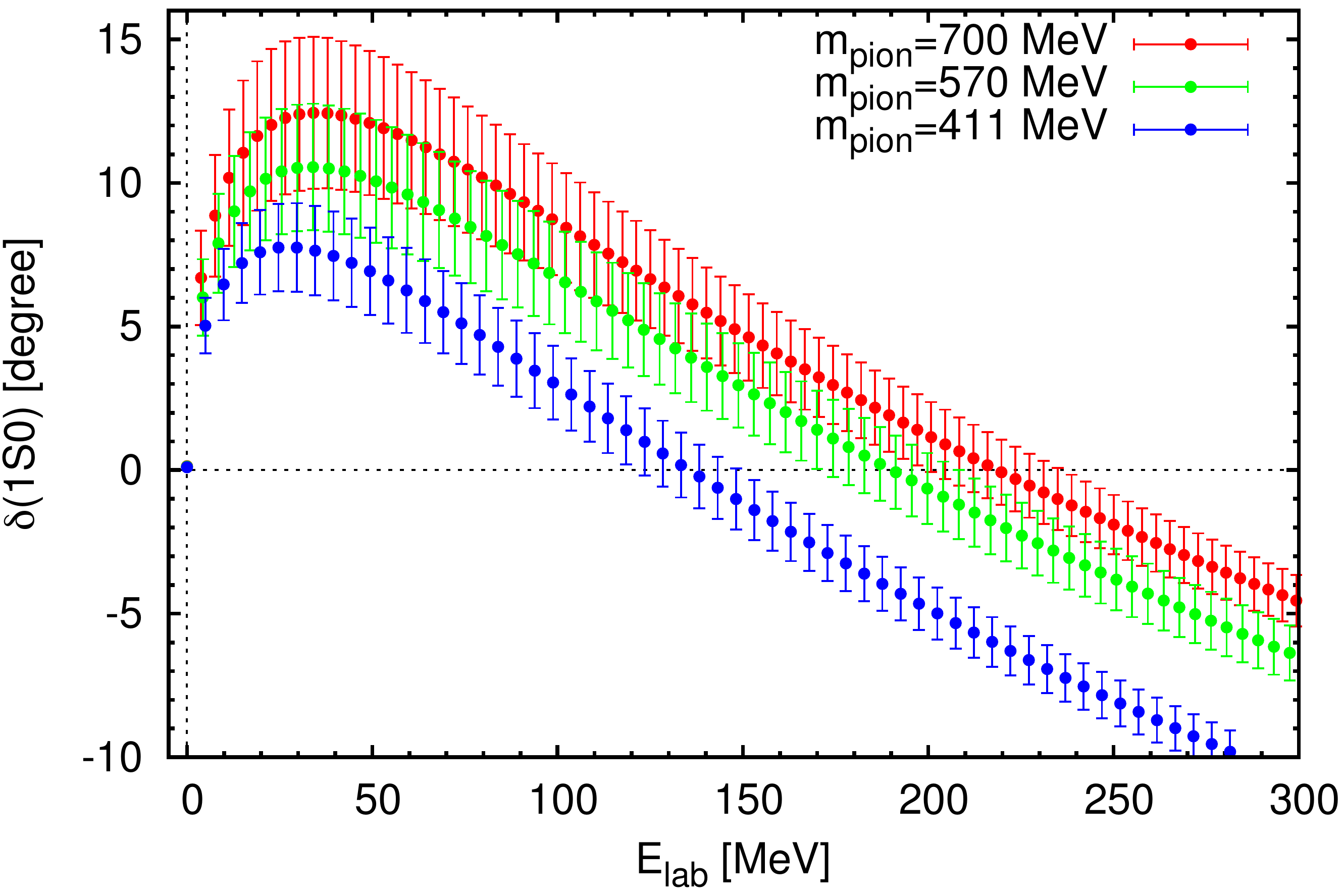}
\\
$m_\pi=800$~MeV& $m_\pi = 138$~MeV& $m_\pi = \{411,570,700\}$~MeV
\end{tabular}
\caption{\label{fig:1s0_phase}The ${}^1 S_0$ phase shift from NPLQCD~\cite{Beane:2013br} (left), experiment (middle) and HALQCD~\cite{Ishii:2013ira} (right).}
\end{figure}

Early lattice calculations of the $NN$ system indicate the scattering lengths relaxed to more natural values for pion masses $m_\pi \gtrsim 350$~MeV~\cite{Beane:2006mx,Aoki:2008hh,Beane:2009py}.
Of particular significance, recent high-statistics calculations have uncovered a bound state in the di-neutron system for heavier pion masses~\cite{Beane:2011iw,Yamazaki:2012hi,Beane:2012vq}.
In particular the results from Refs.~\cite{Yamazaki:2012hi,Beane:2012vq} very clearly show a bound di-neutron.  Example effective mass plots are displayed in Fig.~\ref{fig:nn_bound}.
Levinson's Theorem relates the zero energy phase shift to the number of bound states $\d(0) = n\, \pi/2$.
The barely unbound di-neutron, experimentally, therefore gives rise to a phase shift with a large low-energy peak.
The LQCD calculations with a bound di-neutron will have qualitatively similar phase shifts at larger momentum, but at low energy will noticeably differ.  For example, NPLQCD has computed the phase shift in the $NN$ channels with the ${}^1 S_0$ results displayed in Fig.~\ref{fig:1s0_phase}.
This is to be contrasted with the HALQCD results using the potential methods described in Secs.~\ref{sec:halqcd_1}, \ref{sec:time_dep_pot}, which do not find a bound di-neutron (or deuteron).
While it may seem reassuring that the HALQCD results look qualitatively similar to experiment, if there is a bound di-neutron, their results are not correct at low energies.

\begin{figure}
\center
\begin{tabular}{ccc}
\includegraphics[width=0.315\textwidth]{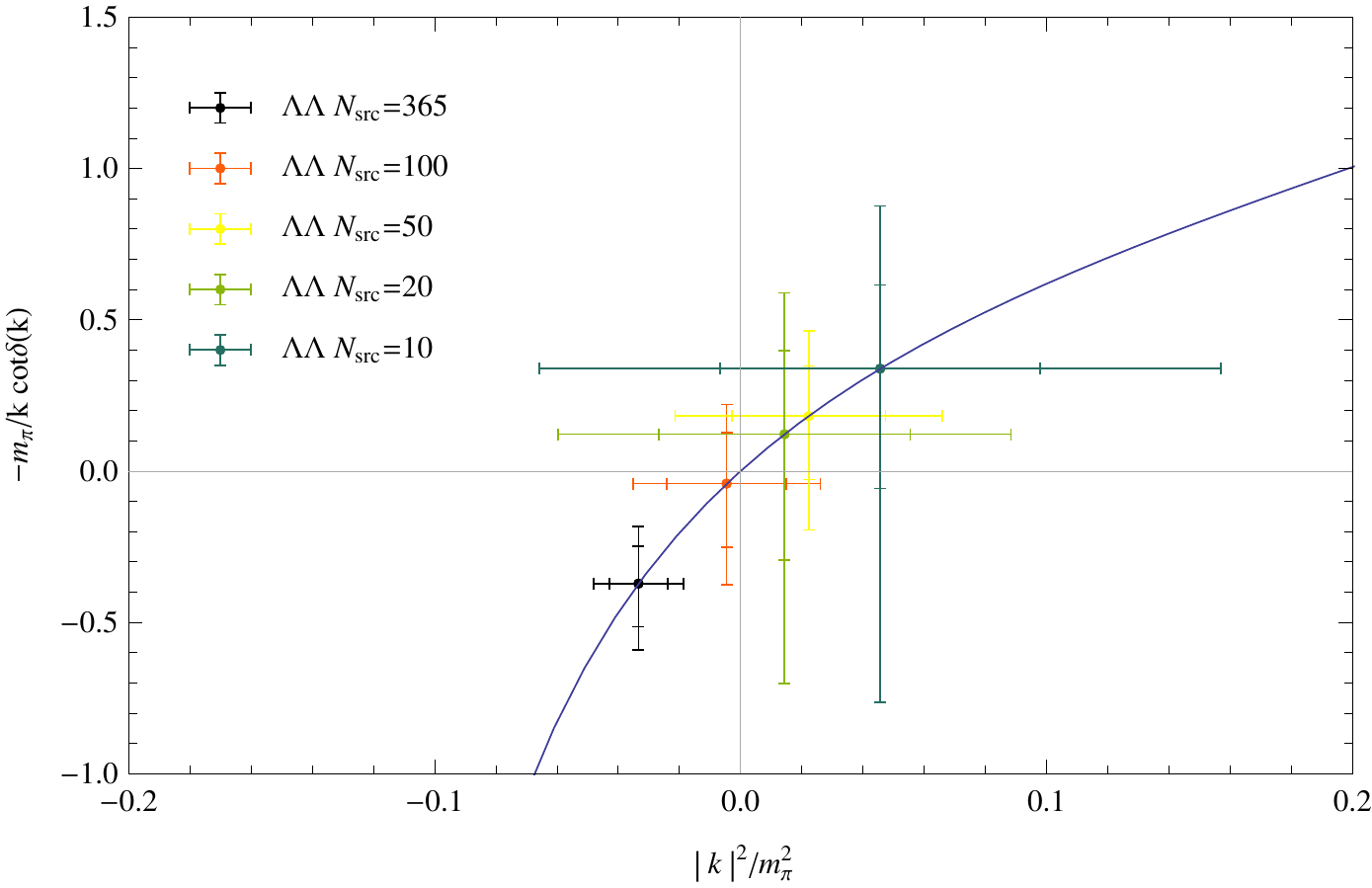}
&\includegraphics[width=0.32\textwidth]{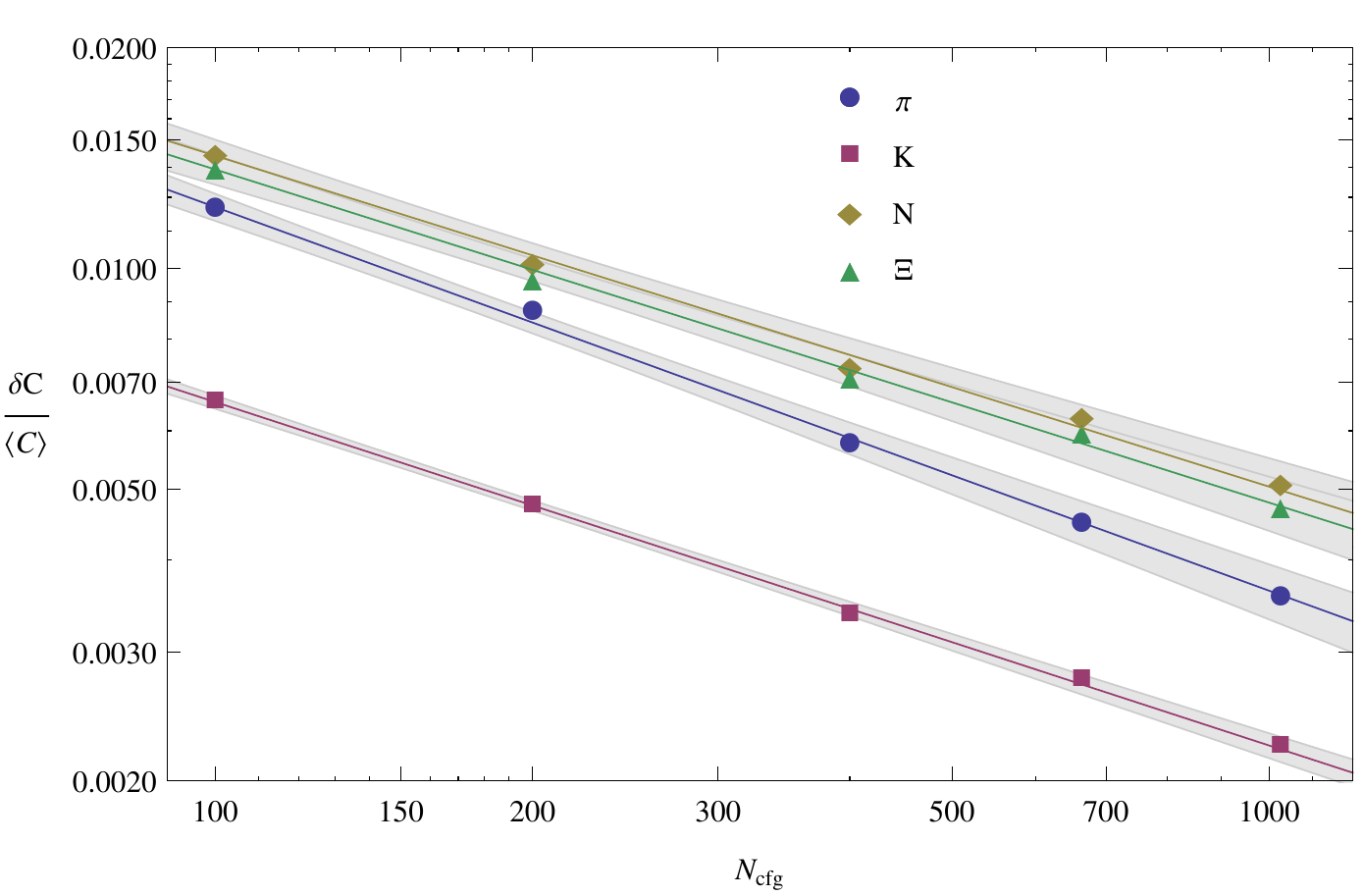}
&\includegraphics[width=0.32\textwidth]{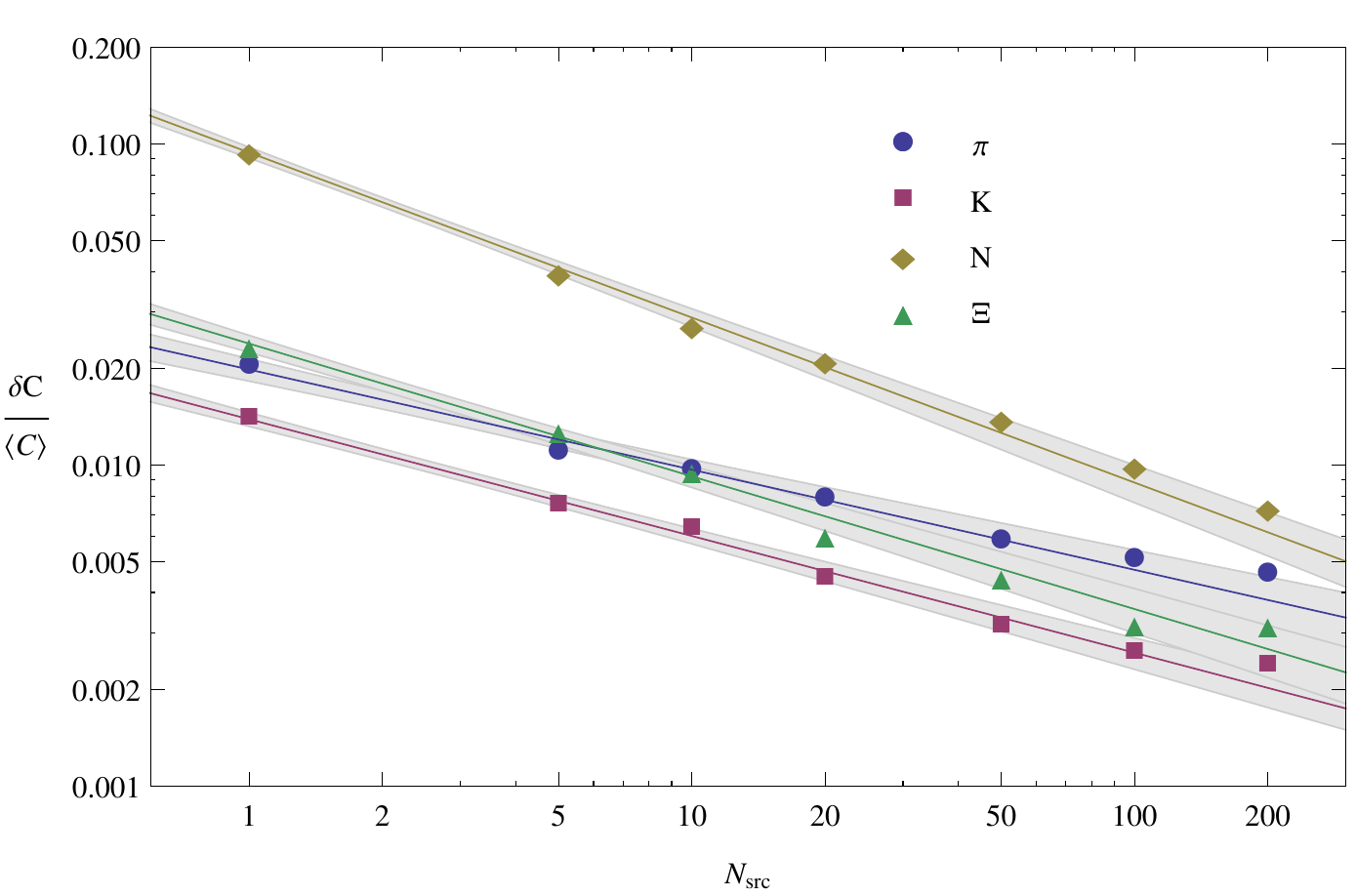}
\\
$k^2_{\L\L}$ versus $N_{src}$ \cite{Beane:2009py}& $\d C_N / <C_N> \propto (N_{cfg})^{-0.45(2)}$& $\d C_N / <C_N> \propto (N_{src})^{-0.51(9)}$
\end{tabular}
\caption{\label{fig:high_stat}NPLQCD study of high statistics calculations \cite{Beane:2009py,Beane:2009kya}.}
\end{figure}

There are now two independent LQCD calculations which find a bound di-neutron at $m_\pi \geq 390$~MeV, NPLQCD~\cite{Beane:2012vq,Beane:2013br} and Yamazaki \textit{et al.}~\cite{Yamazaki:2012hi}.
The results of \cite{Yamazaki:2012hi} use the same gauge-action as HALQCD but are determined with the L\"{u}scher method rather than the potential method.  
\textit{My speculation}: HALQCD does not have enough statistics to resolve the long-range potential, which contributes significantly to the low-energy phase shift, and thus their calculation is unable to resolve the bound di-neutron at these heavier pion masses.
In a previous calculation~\cite{Beane:2009py}, we demonstrated the importance of high statistics for two-baryon systems.
The h-dibaryon binding momentum $k^2$ was computed on $N_{cfg}=1194$ for a number of random sources per configuration ranging from 10 to 365.  The results for all number of ``measurements'' were consistent but only with the highest statistics $N_{meas} = 365\times 1194$ were we able to resolve the h-dibaryon was bound with two-sigma uncertainty.  See Fig.~\ref{fig:high_stat}.
In previous work \cite{Beane:2009kya}, NPLQCD demonstrated that increasing the number of random sources per configuration led to scaling consistent with $\sqrt{N}$ improvement.
This is not surprising since the Compton wavelength of the one and two baryon systems is small compared to the box size.
In contrast, the HALQCD potential method currently only allows for a single source per configuration, and so the statistical precision they have achieved so far is likely insufficient for these calculations.

\subsubsection*{(Hyper)-Nuclei abound}
In order to test new multi-baryon contraction technology~\cite{Detmold:2012eu} NPLQCD has performed high-statistics calculations in the $SU(3)$ flavor limit with $m_{\pi,K} \simeq 800$~MeV~\cite{Beane:2012vq}, 
At these heavy pion masses, it was found that in every two, three and four baryon channel, there were bound states.  In the h-dibaryon channel, there were even two bound states.
Yamazaki \textit{et al.} performed a calculation with $m_\pi\sim510$ and were able to identify bound He nuclei in addition to the deuteron and di-neutron.
They have preliminary results for $m_\pi\sim300$~MeV~\cite{Yamazaki:2013rna} but unsurprisingly conclude significantly more statistics are needed.

\subsection{Challenges and Progress}

\subsubsection{Contractions}
A naive implementation of the Wick contraction necessary for computing multi-baryon systems would lead to a contraction cost that exceeds all of the other costs added up (configuration generation, propagator inversion, ...).
With the more sophisticated operators that will be necessary to properly study these systems, coming up with fast contraction algorithms is essential.
This is a challenge that is being well addressed but the contractions will remain a significant percentage of the computational cost~\cite{Yamazaki:2009ua,Detmold:2012eu,Doi:2012xd,Gunther:2013xj}.

\subsubsection{Finite Volume dependence and boosted systems}

\begin{figure}
\center
\begin{tabular}{cc}
\includegraphics[width=0.35\textwidth]{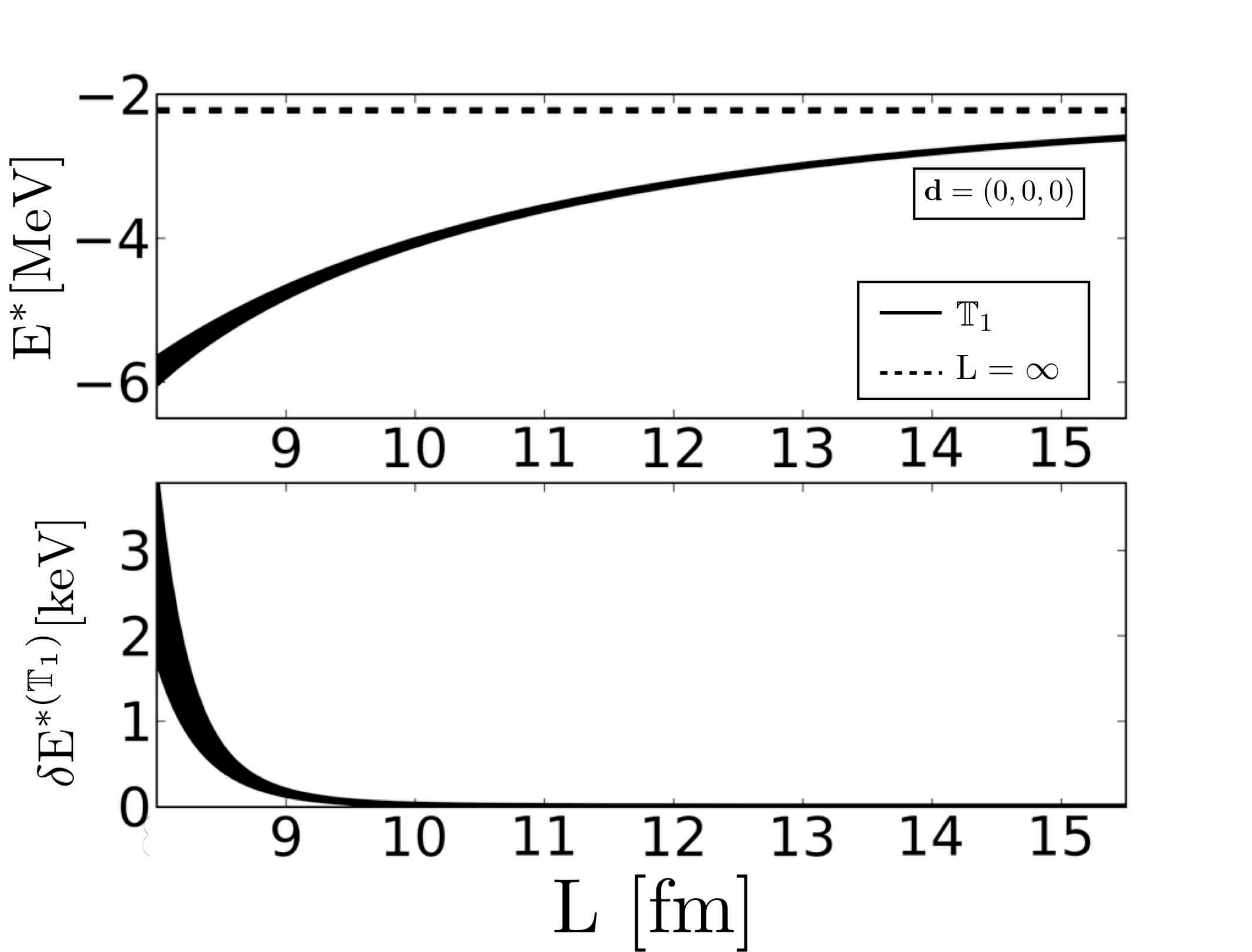}
&\includegraphics[width=0.35\textwidth]{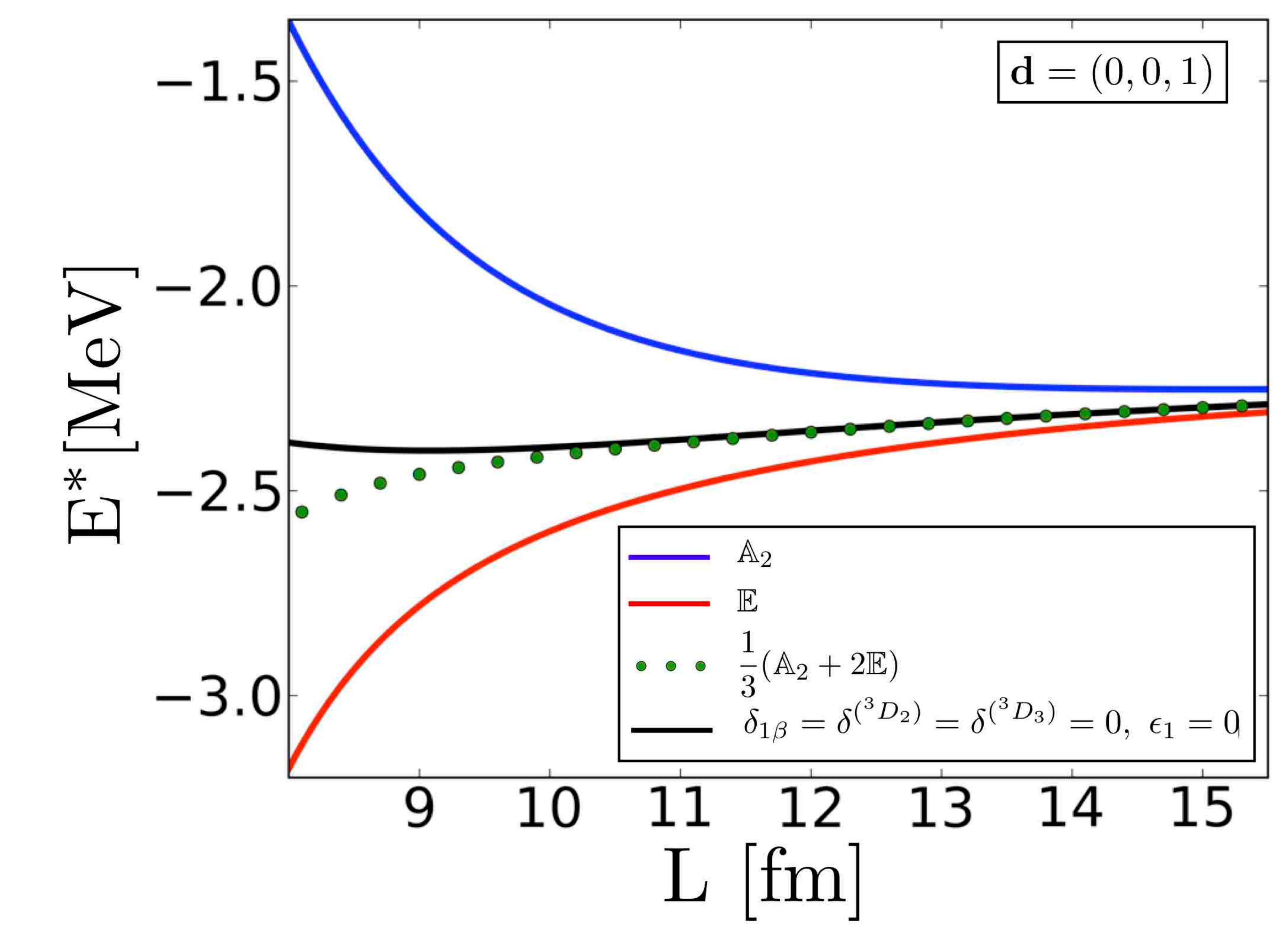}
\end{tabular}
\caption{\label{fig:deut_fv}Estimated deuteron energy levels in finite volume at the physical pion mass including the $S-D$ wave partial mixing from Brice\~{n}o \textit{et al.}~\cite{fv_formalism}.
In the $A_1$ ($T_1$) cubic representation, the energy level is significantly distorted from its infinite volume value and not sensitive to the $S-D$ wave mixing (left).  A projection into the $A_2$ and $E$ representations of the cubic group allows for one to distinguish both the magnitude and sign of the mixing parameter $\bar{\epsilon}$.}
\end{figure}

A detailed understanding of the interactions of two-particles in a finite volume is necessary to determine as much information as possible from the LQCD calculations.  As an example, consider the deuteron.  The deuteron is mostly an $S$-wave with a small $D$-wave admixture.  One way to parameterize the S-matrix in this 2-channel system is
\begin{equation}
S_{2\rightarrow2} = \left( \begin{array}{cc}
	e^{2i \d_1} \cos 2 \bar{\epsilon}& i e^{i(\d_1 + \d_2)} \sin 2 \bar{\epsilon} \\
	i e^{i(\d_1 + \d_2)} \sin 2 \bar{\epsilon}& e^{2i \d_2} \cos 2 \bar{\epsilon}
	\end{array} \right)
\end{equation}
where $\d_{1,2}$ are the phase shifts in the two channels and $\bar{\epsilon}$ parameterizes the mixing between them.
At the physical pion mass, the deuteron energy level is significantly distorted from its infinite volume value until very large volumes are obtained.  It is also insensitive to the $S-D$ wave mixing, see Fig.~\ref{fig:deut_fv} for an estimate of the volume dependence using values of $\d_{1,2}$ and $\bar{\epsilon}$ determined from experiment.
However, the mixing between the $S$ and $D$ wave components of the deuteron can be exploited to determine all the scattering parameters by looking at boosted $NN$ systems projected into the $A_2$ and $E$ representations of the cubic group.  These projections also bring the energy levels closer to the infinite volume value of $B_d \simeq 2.22$~MeV.  See Brice\~{n}o \textit{et al.} for details~\cite{fv_formalism}.

\subsubsection{Coupled Channels and Inelastic States}
\begin{figure}
\center
\begin{tabular}{cc}
\includegraphics[width=0.35\textwidth]{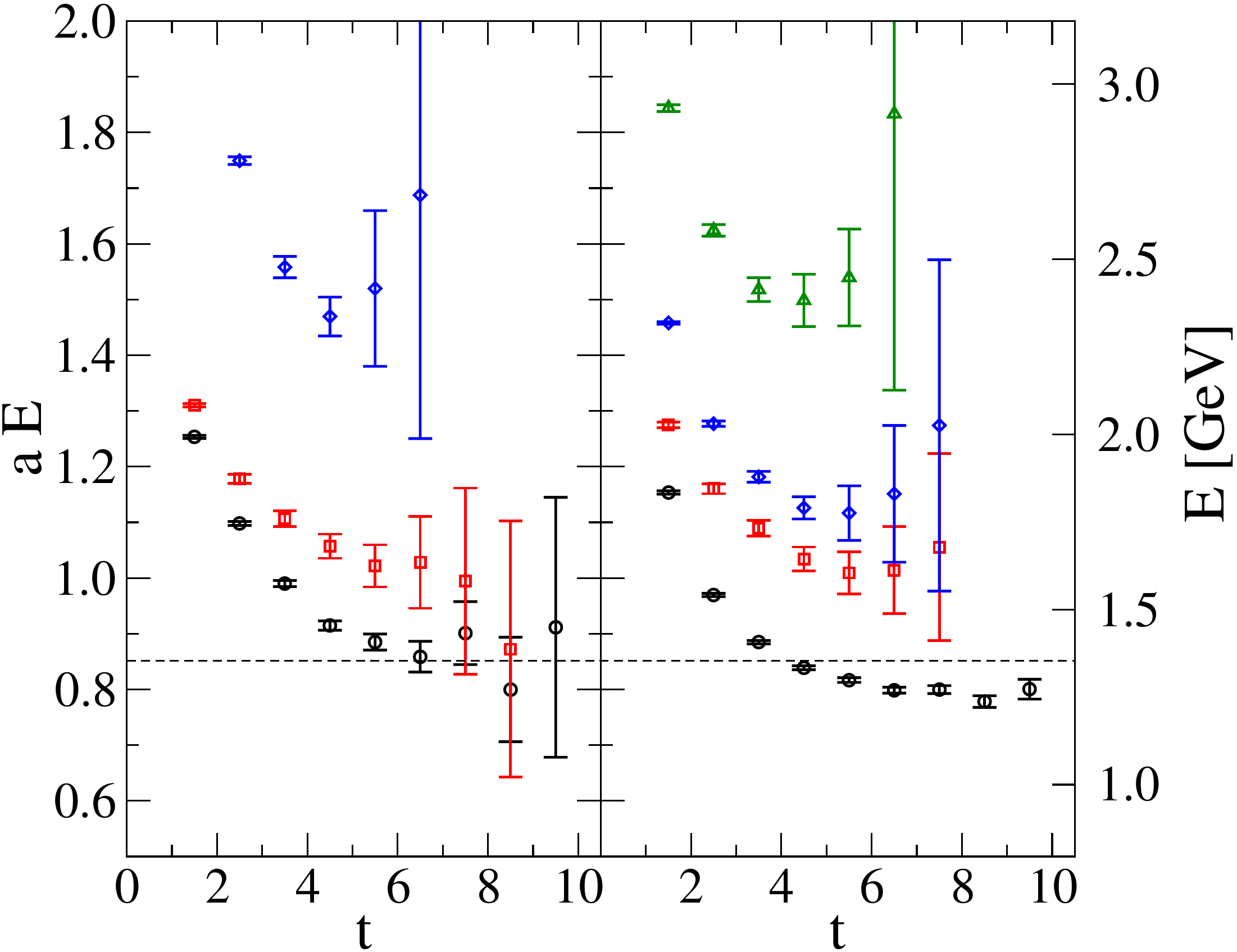}
&\includegraphics[width=0.29\textwidth]{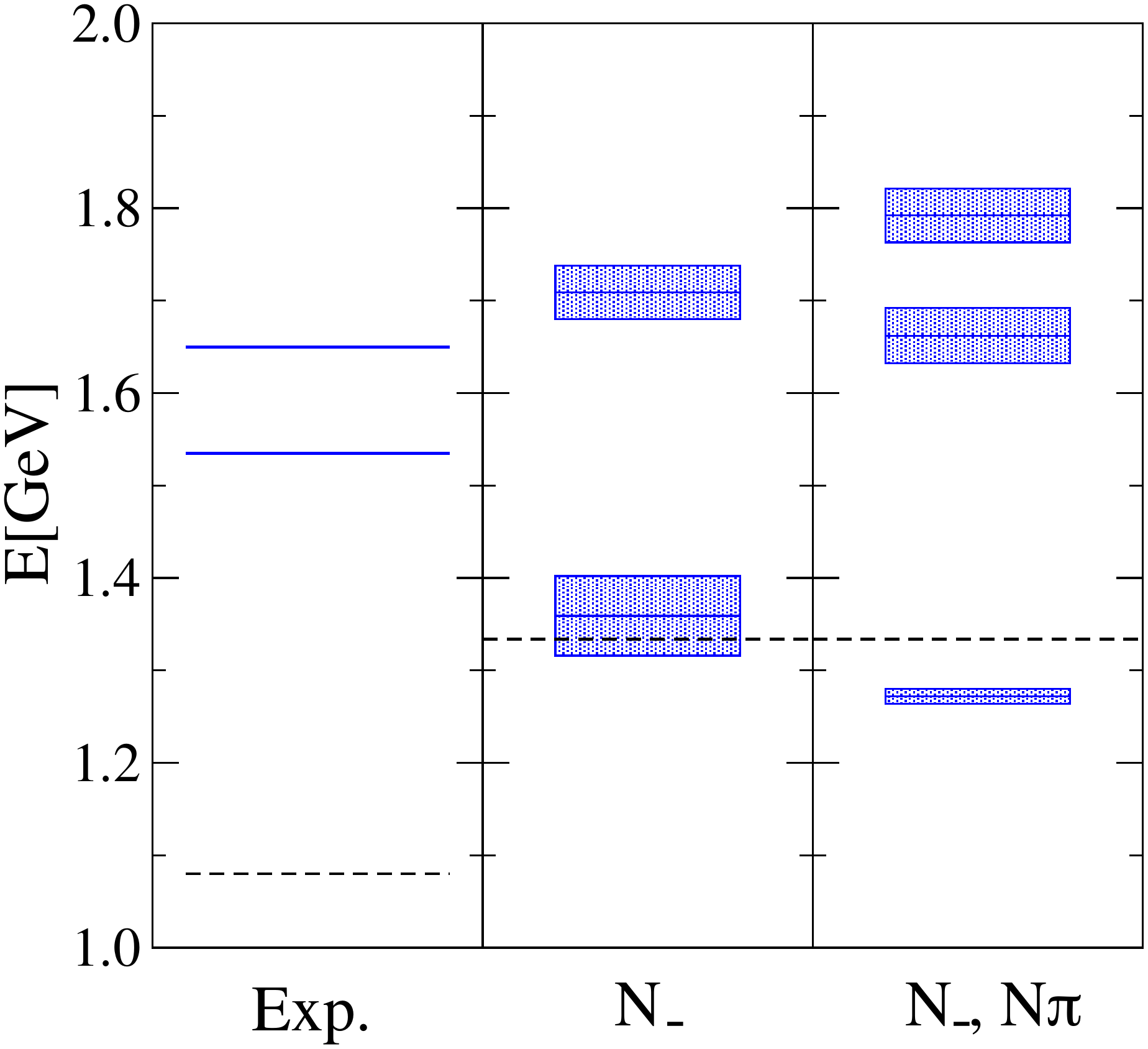}
\end{tabular}
\caption{\label{fig:n_star}A calculation of the negative parity nucleon energy levels using only local operators versus operators which include those resembling $N\pi$ states~\cite{Lang:2012db}.  The use of only local $qqq$ operators leads to a systematically incorrect determination of the QCD eigenstates.  The left figure shows the effective mass plots for only $qqq$ (left-left) and $qqq + N\pi$ (left-right) interpolating fields.
The right plot shows the resulting spectrum from these calculations.}
\end{figure}

One of the most striking observations recently concerns the overlap of various interpolating fields onto the states of interest, and in particular, what happens when one does not use a sufficient basis of operators to couple to all the relevant states, as first demonstrated for $I=1\ \pi\pi$ system near the $\rho$ threshold~\cite{Dudek:2012xn}.
In Ref.~\cite{Lang:2012db}, a calculation of the negative-parity nucleon was performed using i) only local $qqq$ interpolating operators and ii) both $qqq$ operators as well as those which resemble $N\pi$ states.
Neglecting to include the $N\pi$ interpolating fields led to a determination of the spectrum which was systematically incorrect (well outside quoted uncertainties) including the determination of the ground state, See Fig.~\ref{fig:n_star}.
The importance of these observations can not be overstated:
\textit{In order to accurately compute the spectrum, one must include a sufficient basis of interpolating fields to couple to all relevant eigenstates}.
This will become increasingly important for the $NN$ calculations as the pion mass is reduced and large volumes are utilized as the nearby $NN\pi$ states will become increasingly relevant and coupled to the $NN$ states.

For two coupled channels, there are at least three pieces of information which must be computed at the same energy to determine the two phase shifts and the mixing parameter (at that energy).
But for a fixed volume and total momentum, a LQCD calculation will determine these pieces of information at different energies.
What is needed is a means of smoothly parameterizing the phase shift.
If the HALQCD potential method can be demonstrated to be consistent with the L\"{u}scher method for interesting systems, it has the potential to be the perfect tool for this parameterization.
This is particularly true for coupled channels which are related by a symmetry that can be expressed at the level of interpolating fields, such as the $I=0$ $\{\pi\pi,KK\}$ system with $4m_\pi > 2m_K$.
The approximate chiral symmetry allows one to fix the normalization between the interpolating fields used, up to a desired power in $SU(3)$ breaking corrections, to couple to the various channels and thus determine the strength of the mixing potential with respect to the $\pi\pi$ and $KK$ potentials.
Otherwise, the relative strength of the mixing potential may depend upon the choice of interpolating field through the numerical determination of the potential, Eq.~\eqref{eq:halqcd_pot}.
For a single channel, the overall normalization does not matter.  But for the $NN$ and $NN\pi$ coupled system, the choice of interpolating field may pollute the mixing potential.

\subsubsection{Three particles in a box}
One of the next steps is to understand the quantization conditions for three particles in a finite volume.
I refer the interested reader to the brief literature on the subject (so far)~\cite{Beane:2007es,Beane:2009gs,three_particles}.

\subsubsection{Other interesting multi-hadron calculations}
There are many other interesting areas of research in multi-hadron physics which I did not have time to even mention: baryon properties in a sea of mesons~\cite{Detmold:2013gua};
lattice discretized $NN$ EFT which has allowed for calculations of up to ${}^{16}$O, reviewed in the talk at this conference~\cite{Lahde:2013kma},
and very many other interesting talks and publications.

\subsection{Conclusions}
\begin{itemize}
\item Nuclear physics is beginning a renaissance with lattice QCD and EFT providing the tools to connect low-energy nuclear physics with the fundamental theory of strong interactions

\item It is currently a very open field with room and need for new ideas

\item It is exciting to see more people and groups getting involved, especially so many young scientists who are driving new developments.

\item There are significant challenges which need to be overcome.  In my opinion, the most important is the issue of the basis of interpolating fields used.  In order to resolve the disparate energy scales relevant in multi-baryon correlation functions, very good projections must be made onto all the relevant states.  Without a good and complete basis, we see the results will be systematically incorrect, as demonstrated with the negative parity nucleon~\cite{Lang:2012db}.
This will also be important for dealing with the nearly degenerate inelastic channels, such as $NN \rightarrow NN\pi$ which will become accessible for light pion masses in large boxes.

\item The formalism for three particles in a finite box will soon be applied to numerical results.  This is the first step to computing for example, the three-neutron interactions necessary to understand larger nuclei and nuclear matter.

\item It will be interesting to see if the HALQCD potential method, augmented with perhaps a variational basis of interpolating fields, can resolve a bound state in the $NN$ channels.

\end{itemize}

\section*{Acknowledgements}
\noindent
I would like to particularly thank Ra\'{u}l Brice\~{n}o and Max Hansen for many detailed conversations.  I would also like to thank Bruno Charron, Takumi Doi and Tetsuo Hatsuda for many fruitful discussions.  I would like to thank all the members of HALQCD for generously sharing with me many details of their work, as well as intermediate numerical results which was very helpful in preparing this review.  I would also like to thank everyone else who provided material for this review and Ra\'{u}l Brice\~{n}o and Will Detmold for comments on these proceedings.


\begin{thebibliography}{99}
\bibitem{kw_award}
  \href{http://www.lattice2013.uni-mainz.de/219_ENG_HTML.php}{2013 Kenneth G. Wilson Award},
  \href{http://www.lattice2013.uni-mainz.de/219_ENG_HTML.php}{http://www.lattice2013.uni-mainz.de/219\_ENG\_HTML.php}.
%
\bibitem{Kronfeld:2013lda}
  A.~S.~Kronfeld,
  PoS LATTICE {\bf 2013} (2013) 504
  [arXiv:1312.6861 [physics.hist-ph]].
\bibitem{WalkerLoud:2008bp}
  A.~Walker-Loud \textit{et al.} [LHP Collaboration],
  Phys.\ Rev.\ D {\bf 79} (2009) 054502
  [arXiv:0806.4549].
\bibitem{WalkerLoud:2008pj}
  A.~Walker-Loud,
  PoS LATTICE {\bf 2008} (2008) 005
  [arXiv:0810.0663 [hep-lat]].
\bibitem{Walker-Loud:2013yua}
  A.~Walker-Loud,
  PoS CD {\bf 12} (2013) 017
  [arXiv:1304.6341 [hep-lat]].
\bibitem{Gong:2013vja}
  M.~Gong \textit{et al.} [$\chi$QCD Collaboration],
  Phys.\ Rev.\ D {\bf 88} (2013) 014503
  [arXiv:1304.1194 [hep-ph]].
\bibitem{Young:2013nn}
  R.~D.~Young,
  PoS LATTICE {\bf 2012} (2012) 014
  [arXiv:1301.1765 [hep-lat]].
\bibitem{dm}
  A.~Bottino \textit{et al.}, 
  Astropart.\ Phys.\  {\bf 13} (2000) 215
  [hep-ph/9909228];
  Astropart.\ Phys.\  {\bf 18} (2002) 205
  [hep-ph/0111229];
  J.~R.~Ellis \textit{et al.},
  Phys.\ Rev.\ D {\bf 77} (2008) 065026
  [arXiv:0801.3656 [hep-ph]];
  R.~J.~Hill and M.~P.~Solon,
  Phys.\ Lett.\ B {\bf 707} (2012) 539
  [arXiv:1111.0016];
  arXiv:1309.4092;
  C.~Cheung, L.~J.~Hall, D.~Pinner and J.~T.~Ruderman,
  JHEP {\bf 1305} (2013) 100
  [arXiv:1211.4873].

\bibitem{Mohr:2008fa}
  P.~J.~Mohr, B.~N.~Taylor and D.~B.~Newell,
  Rev.\ Mod.\ Phys.\  {\bf 80} (2008) 633
  [arXiv:0801.0028].
\bibitem{Griffiths:2008zz}
  D.~Griffiths,
  Weinheim, Germany: Wiley-VCH (2008) 454 p
\bibitem{Dunkley:2008ie}
  J.~Dunkley {\it et al.}  [WMAP Collaboration],
  Astrophys.\ J.\ Suppl.\  {\bf 180} (2009) 306
  [arXiv:0803.0586].
\bibitem{Olive:2010mh}
  K.~A.~Olive,
  CERN Yellow Report CERN-2010-002, 149-196
  [arXiv:1005.3955 [hep-ph]].

\bibitem{lattice_isospin}
  B.~C.~Tiburzi and A.~Walker-Loud,
  Nucl.\ Phys.\ A {\bf 764} (2006) 274
  [hep-lat/0501018];
  S.~R.~Beane \textit{et al.}, 
  Nucl.\ Phys.\ B {\bf 768} (2007) 38
  [hep-lat/0605014];
  A.~Walker-Loud,
  arXiv:0904.2404;
  T.~Blum \textit{et al.}, 
  Phys.\ Rev.\ D {\bf 82} (2010) 094508
  [arXiv:1006.1311];
  G.~M.~de Divitiis \textit{et al.} [RM123], 
  JHEP {\bf 1204} (2012) 124
  [arXiv:1110.6294];
  R.~Horsley {\it et al.}  [QCDSF and UKQCD],
  Phys.\ Rev.\ D {\bf 86} (2012) 114511
  [arXiv:1206.3156];
  G.~M.~de Divitiis \textit{et al.} [RM123], 
  Phys.\ Rev.\ D {\bf 87} (2013) 114505
  [arXiv:1303.4896];
  S.~Borsanyi \textit{et al.} [BMW], 
  Phys.\ Rev.\ Lett.\  {\bf 111} (2013) 252001
  [arXiv:1306.2287].


\bibitem{Cottingham} 
  M.~Cini, E.~Ferrari and R.~Gatto,
  Phys.~Rev.~Lett.~\textbf{2}, 7 (1959);
  W.~N.~Cottingham,
  Annals Phys.\  {\bf 25}, 424 (1963);
  J.~Gasser and H.~Leutwyler,
  Nucl.\ Phys.\ B {\bf 94}, 269 (1975);
  Phys.\ Rept.\  {\bf 87}, 77 (1982);
  J.~C.~Collins,
  Nucl.\ Phys.\ B {\bf 149}, 90 (1979)
  [Erratum-ibid.\ B {\bf 153}, 546 (1979)].
\bibitem{WalkerLoud:2012bg}
  A.~Walker-Loud, C.~E.~Carlson and G.~A.~Miller,
  Phys.\ Rev.\ Lett.\  {\bf 108} (2012) 232301
  [arXiv:1203.0254];
  PoS LATTICE {\bf 2012} (2012) 136
  [arXiv:1210.7777].
\bibitem{Griesshammer:2012we}
  H.~W.~Griesshammer \textit{et al.}, 
  Prog.\ Part.\ Nucl.\ Phys.\  {\bf 67} (2012) 841
  [arXiv:1203.6834].
\bibitem{emc:pols}
  W.~Detmold, B.~C.~Tiburzi and A.~Walker-Loud,
  Phys.\ Rev.\ D {\bf 73} (2006) 114505
  [hep-lat/0603026];
  Phys.\ Rev.\ D {\bf 79} (2009) 094505
  [arXiv:0904.1586];
  Phys.\ Rev.\ D {\bf 81} (2010) 054502
  [arXiv:1001.1131];
  B.~C.~Tiburzi,
  Nucl.\ Phys.\ A {\bf 814} (2008) 74
  [arXiv:0808.3965].

  
\bibitem{Aoki:2013ldr}
  S.~Aoki \textit{et al.}, [FLAG Working Group], 
  arXiv:1310.8555.

\bibitem{Junnarkar:2013ac}
  P.~Junnarkar and A.~Walker-Loud,
  Phys.\ Rev.\ D {\bf 87} (2013) 114510
  [arXiv:1301.1114 [hep-lat]].
\bibitem{WalkerLoud:2009nf}
  A.~Walker-Loud,
  arXiv:0904.2404 [hep-lat].

\bibitem{Sharpe:1997by}
  S.~R.~Sharpe,
  Phys.\ Rev.\ D {\bf 56} (1997) 7052
   [Erratum-ibid.\ D {\bf 62} (2000) 099901]
  [hep-lat/9707018].

\bibitem{Lin:2008pr}
  H.-W.~Lin {\it et al.}  [Hadron Spectrum Coll.],
  Phys.\ Rev.\ D {\bf 79} (2009) 034502
  [arXiv:0810.3588].

\bibitem{emc_spec}
  C.~Aubin, W.~Detmold, E.~Mereghetti, K.~Orginos, S.~Syritsyn, B.~Tiburzi and A.~Walker-Loud [ElectroMagnetic Collaboration],
  \textit{in progress}.

\bibitem{WalkerLoud:2011ab}
  A.~Walker-Loud,
  Phys.\ Rev.\ D {\bf 86} (2012) 074509
  [arXiv:1112.2658 [hep-lat]].

\bibitem{bbn_md-mu}
  P.~Banerjee, T.~C.~Luu and A.~Walker-Loud,
  \textit{in preparation}

%
\bibitem{Maiani:1990ca}
  L.~Maiani and M.~Testa,
  Phys.\ Lett.\ B {\bf 245} (1990) 585.
\bibitem{Luscher:1986pf}
  M.~Luscher,
  Commun.\ Math.\ Phys.\  {\bf 105} (1986) 153;
  Nucl.\ Phys.\ B {\bf 354} (1991) 531.
\bibitem{Bedaque:2006yi}
  P.~F.~Bedaque, I.~Sato and A.~Walker-Loud,
  Phys.\ Rev.\ D {\bf 73} (2006) 074501
  [hep-lat/0601033];
  I.~Sato and P.~F.~Bedaque,
  Phys.\ Rev.\ D {\bf 76} (2007) 034502
  [hep-lat/0702021].

\bibitem{fv_formalism}
  K.~Rummukainen and S.~A.~Gottlieb,
  Nucl.\ Phys.\ B {\bf 450} (1995) 397
  [hep-lat/9503028];
  C.~H.~Kim \textit{et al.}, 
  Nucl.\ Phys.\ B {\bf 727} (2005) 218
  [hep-lat/0507006];
  C.~Liu \textit{et al.}, 
  Int.\ J.\ Mod.\ Phys.\ A {\bf 21} (2006) 847
  [hep-lat/0508022];
  N.~Ishizuka,
  PoS LAT {\bf 2009} (2009) 119
  [arXiv:0910.2772];
  T.~Luu and M.~J.~Savage,
  Phys.\ Rev.\ D {\bf 83} (2011) 114508
  [arXiv:1101.3347];
  M.~T.~Hansen and S.~R.~Sharpe,
  Phys.\ Rev.\ D {\bf 86} (2012) 016007
  [arXiv:1204.0826];
  R.~A.~Briceno and Z.~Davoudi,
  Phys.\  Rev.\  D.\  88, {\bf 094507} (2013)
   [Phys.\ Rev.\ D {\bf 88} (2013) 094507]
  [arXiv:1204.1110];
  R.~A.~Briceno \textit{et al}, 
  Phys.\ Rev.\ D {\bf 88} (2013) 034502
  [arXiv:1305.4903];
  R.~A.~Briceno \textit{et al}, 
  Phys.\ Rev.\ D {\bf 88} (2013) 114507
  [arXiv:1309.3556];
  arXiv:1311.7686;
  R.~A.~Briceno,
  arXiv:1401.3312;
  N.~Li \textit{et al.}, 
  arXiv:1401.5569.
  
\bibitem{Beane:2011sc}
  S.~R.~Beane {\it et al.}  [NPLQCD Collaboration],
  Phys.\ Rev.\ D {\bf 85} (2012) 034505
  [arXiv:1107.5023].
\bibitem{Dudek:2012gj}
  J.~J.~Dudek, R.~G.~Edwards and C.~E.~Thomas,
  Phys.\ Rev.\ D {\bf 86} (2012) 034031
  [arXiv:1203.6041].

\bibitem{Ishii:2006ec}
  N.~Ishii, S.~Aoki and T.~Hatsuda,
  Phys.\ Rev.\ Lett.\  {\bf 99} (2007) 022001
  [nucl-th/0611096].
\bibitem{Aoki:2005uf}
  S.~Aoki {\it et al.}  [CP-PACS Collaboration],
  Phys.\ Rev.\ D {\bf 71} (2005) 094504
  [hep-lat/0503025].
\bibitem{HALQCD:2012aa}
  N.~Ishii {\it et al.}  [HAL QCD Collaboration],
  Phys.\ Lett.\ B {\bf 712} (2012) 437
  [arXiv:1203.3642 [hep-lat]].
\bibitem{Kurth:2013tua}
  T.~Kurth, N.~Ishii, T.~Doi, S.~Aoki and T.~Hatsuda,
  JHEP {\bf 1312} (2013) 015
  [arXiv:1305.4462].
\bibitem{Charron:2013paa}
  B.~Charron [HAL QCD Collaboration],
  arXiv:1312.1032 [hep-lat].
\bibitem{Luscher:1990ck}
  M.~Luscher and U.~Wolff,
  Nucl.\ Phys.\ B {\bf 339} (1990) 222.
\bibitem{Chen:2005ab}
  J.~-W.~Chen, D.~O'Connell, R.~S.~Van de Water and A.~Walker-Loud,
  Phys.\ Rev.\ D {\bf 73} (2006) 074510.

\bibitem{Beane:2006mx}
  S.~R.~Beane \textit{et al.} [NPLQCD Collaboration],
  Phys.\ Rev.\ Lett.\  {\bf 97} (2006) 012001
  [hep-lat/0602010].
\bibitem{Fukugita:1994ve}
  M.~Fukugita, Y.~Kuramashi, M.~Okawa, H.~Mino and A.~Ukawa,
  Phys.\ Rev.\ D {\bf 52} (1995) 3003.
\bibitem{Beane:2007es}
  S.~R.~Beane \textit{et al.}, 
  Phys.\ Rev.\ Lett.\  {\bf 100} (2008) 082004
  [arXiv:0710.1827];
  Phys.\ Rev.\ D {\bf 78} (2008) 014507
  [arXiv:0803.2728];
W.~Detmold \textit{et al.}, 
  Phys.\ Rev.\ D {\bf 78} (2008) 054514
  [arXiv:0807.1856];
  Phys.\ Rev.\ Lett.\  {\bf 102} (2009) 032004
  [arXiv:0809.0892].

\bibitem{Beane:2009gs}
  S.~R.~Beane \textit{et al.} [NPLQCD Collaboration],
  Phys.\ Rev.\ D {\bf 80} (2009) 074501
  [arXiv:0905.0466].

\bibitem{Yamazaki:2009ua}
  T.~Yamazaki {\it et al.}  [PACS-CS Collaboration],
  Phys.\ Rev.\ D {\bf 81} (2010) 111504
  [arXiv:0912.1383].
\bibitem{Beane:2010hg}
  S.~R.~Beane {\it et al.}  [NPLQCD Collaboration],
  Phys.\ Rev.\ Lett.\  {\bf 106} (2011) 162001
  [arXiv:1012.3812].
\bibitem{Inoue:2010es}
  T.~Inoue {\it et al.}  [HAL QCD Collaboration],
  Phys.\ Rev.\ Lett.\  {\bf 106} (2011) 162002
  [arXiv:1012.5928].
\bibitem{Jaffe:1976yi}
  R.~L.~Jaffe,
  Phys.\ Rev.\ Lett.\  {\bf 38} (1977) 195
   [Erratum-ibid.\  {\bf 38} (1977) 617].
\bibitem{h:experiment}
  H.~Takahashi \textit{et al.}, 
  Phys.\ Rev.\ Lett.\  {\bf 87} (2001) 212502.
  C.~J.~Yoon \textit{et al.}, 
  Phys.\ Rev.\ C {\bf 75} (2007) 022201;
  N.~Shah [STAR Collaboration],
  Nucl.\ Phys.\ A904-905 {\bf 2013} (2013) 443c
  [arXiv:1210.5436];
  A.~Ohnishi {\it et al.}  [ExHIC Collaboration],
  Nucl.\ Phys.\ A {\bf 914} (2013) 377
  [arXiv:1301.7261].
  B.~H.~Kim {\it et al.}  [Belle Collaboration],
near $2m_\Lambda$ in $\Upsilon(1S)$ and $\Upsilon(2S)$ decays,''
  Phys.\ Rev.\ Lett.\  {\bf 110} (2013) 222002
  [arXiv:1302.4028].


\bibitem{Beane:2011zpa}
  S.~R.~Beane {\it et al.} [NPLQCD Collaboration],
  Mod.\ Phys.\ Lett.\ A {\bf 26} (2011) 2587
  [arXiv:1103.2821].
\bibitem{Haidenbauer:2011ah}
  J.~Haidenbauer and U.~-G.~Meissner,
  Phys.\ Lett.\ B {\bf 706} (2011) 100
  [arXiv:1109.3590 [hep-ph]].
\bibitem{Francis:2013lva}
  A.~Francis, C.~Miao, T.~D.~Rae and H.~Wittig,
  arXiv:1311.3933 [hep-lat].
\bibitem{Aoki:2008hh}
  S.~Aoki, T.~Hatsuda and N.~Ishii,
  Comput.\ Sci.\ Dis.\  {\bf 1} (2008) 015009
  [arXiv:0805.2462 [hep-ph]].

\bibitem{Beane:2011iw}
  S.~R.~Beane {\it et al.}  [NPLQCD Collaboration],
  Phys.\ Rev.\ D {\bf 85} (2012) 054511
  [arXiv:1109.2889].
\bibitem{Yamazaki:2012hi}
  T.~Yamazaki, K.~Ishikawa, Y.~Kuramashi and A.~Ukawa,
  Phys.\ Rev.\ D {\bf 86} (2012)
  [arXiv:1207.4277].
\bibitem{Beane:2012vq}
  S.~R.~Beane {\it et al.} [NPLQCD Collaboration],
  Phys.\ Rev.\ D {\bf 87} (2013) 3,  034506
  [arXiv:1206.5219].
\bibitem{Beane:2013br}
  S.~R.~Beane {\it et al.} [NPLQCD Collaboration],
  Phys.\ Rev.\ C {\bf 88} (2013) 024003
  [arXiv:1301.5790].
\bibitem{Ishii:2013ira}
  N.~Ishii [HAL QCD Collaboration],
  PoS CD {\bf 12} (2013) 025.
\bibitem{Beane:2009py}
  S.~R.~Beane {\it et al.}  [NPLQCD Collaboration],
  Phys.\ Rev.\ D {\bf 81} (2010) 054505
  [arXiv:0912.4243].
\bibitem{Beane:2009kya}
  S.~R.~Beane {\it et al.}  [NPLQCD Collaboration],
  Phys.\ Rev.\ D {\bf 79} (2009) 114502
  [arXiv:0903.2990].
\bibitem{Detmold:2012eu}
  W.~Detmold and K.~Orginos,
  Phys.\ Rev.\ D {\bf 87} (2013) 114512
  [arXiv:1207.1452 [hep-lat]].
\bibitem{Yamazaki:2013rna}
  T.~Yamazaki, K.~-I.~Ishikawa, Y.~Kuramashi and A.~Ukawa,
  arXiv:1310.5797 [hep-lat].
\bibitem{Doi:2012xd}
  T.~Doi and M.~G.~Endres,
  Comput.\  Phys.\  Commun.\  {\bf 184} (2013) 117
  [arXiv:1205.0585 [hep-lat]].
\bibitem{Gunther:2013xj}
  J.~Gunther, B.~C.~Toth and L.~Varnhorst,
  Phys.\ Rev.\ D {\bf 87} (2013) 094513
  [arXiv:1301.4895 [hep-lat]].

\bibitem{Dudek:2012xn}
  J.~J.~Dudek, R.~G.~Edwards and C.~E.~Thomas,
  Phys.\ Rev.\ D {\bf 87} (2013) 3,  034505
  [arXiv:1212.0830].
\bibitem{Lang:2012db}
  C.~B.~Lang and V.~Verduci,
  Phys.\ Rev.\ D {\bf 87} (2013) 5,  054502
  [arXiv:1212.5055].

\bibitem{three_particles}
  T.~Luu,
  PoS LATTICE {\bf 2008} (2008) 246
  [arXiv:0810.2331];
  S.~Kreuzer and H.-W.~Hammer,
  Phys.\ Lett.\ B {\bf 673} (2009) 260
  [arXiv:0811.0159];
  Eur.\ Phys.\ J.\ A {\bf 43} (2010) 229
  [arXiv:0910.2191];
  Phys.\ Lett.\ B {\bf 694} (2011) 424
  [arXiv:1008.4499];
  K.~Polejaeva and A.~Rusetsky,
  Eur.\ Phys.\ J.\ A {\bf 48} (2012) 67
  [arXiv:1203.1241];
  S.~Kreuzer and H.~W.~Grie{\ss}hammer,
  Eur.\ Phys.\ J.\ A {\bf 48} (2012) 93
  [arXiv:1205.0277];
  R.~A.~Briceno and Z.~Davoudi,
  Phys.\ Rev.\ D {\bf 87} (2013) 094507
  [arXiv:1212.3398];
  M.~T.~Hansen and S.~R.~Sharpe,
  arXiv:1311.4848;



\bibitem{Detmold:2013gua}
  W.~Detmold and A.~N.~Nicholson,
  Phys.\ Rev.\ D {\bf 88} (2013) 074501
  [arXiv:1308.5186 [hep-lat]].
\bibitem{Lahde:2013kma}
  T.~A.~Lähde, E.~Epelbaum, H.~Krebs, D.~Lee, U.~-G.~Meißner and G.~Rupak,
  arXiv:1311.1968.
  
\end{thebibliography}
\end{document}